\documentclass[a4paper,12pt]{article}

\usepackage{amsmath,amsfonts,amssymb}
\usepackage{bm}
\usepackage{physics}
\usepackage{tensor}
\usepackage{ascmac}
\usepackage{here}
\usepackage{xcolor}
\usepackage{graphicx}
\usepackage{bmpsize}
\graphicspath{{image/}}

\usepackage{mathrsfs}
\usepackage{authblk}
\usepackage{title}
\numberwithin{equation}{section}
\definecolor{DarkBlue}{rgb}{0,0,0.9} 

\usepackage[top=30truemm,bottom=25truemm,left=25truemm,right=30truemm]{geometry}
\usepackage[colorlinks=true
,urlcolor=blue
,anchorcolor=blue
,citecolor=blue
,filecolor=blue
,linkcolor=blue
,menucolor=blue
,pagecolor=blue
,linktocpage=true
,pdfproducer=medialab
,pdfa=true
]{hyperref}

\usepackage{cite}

\begin{document}

\title{Dynamical Formation of Charged Wormholes}
\author[$a$,$b$]{\textbf{Yasutaka Koga}\thanks{\href{mailto:yasutaka.koga@oit.ac.jp}{yasutaka.koga@oit.ac.jp}}}
\author[$b$]{\textbf{Ryota Maeda} \thanks{\href{mailto:ryota.maeda@yukawa.kyoto-u.ac.jp}{ryota.maeda@yukawa.kyoto-u.ac.jp}}}
\author[$c$]{\textbf{Daiki 
Saito}
\thanks{\href{mailto:saito@tap.scphys.kyoto-u.ac.jp}{saito@tap.scphys.kyoto-u.ac.jp}}}
\author[$d$]{\textbf{Keiya Uemichi}
\thanks{\href{mailto:uemichi.keiya.j4@s.mail.nagoya-u.ac.jp}{uemichi.keiya.j4@s.mail.nagoya-u.ac.jp}}}

\author[$e$]{\textbf{Daisuke Yoshida}
\thanks{\href{mailto:dyoshida@math.nagoya-u.ac.jp}{dyoshida@math.nagoya-u.ac.jp}}}

\affil[$a$]{{\small	 Department of Information and Computer Science, Osaka Institute of Technology, Hirakata 573-0196, Japan}}
\affil[$b$]{{\small	Center for Gravitational Physics and Quantum Information, Yukawa Institute for Theoretical Physics, Kyoto University\\
	Kitashirakawa Oiwakecho, Sakyo-ku, Kyoto 606-8502, Japan}}
\affil[$c$]{{\small	 Department of Physics, Kyoto University, Kyoto 606-8502, Japan}}
\affil[$d$]{{\small	 Division of Science, Graduate School of Science, Nagoya University, Nagoya 464-8602, Japan}}
\affil[$e$]{{\small	 Department of Mathematics, Nagoya University, Nagoya 464-8602, Japan}}

\begin{flushright}
KUNS-3056, YITP-25-85, NU-QG-7
\end{flushright}

\maketitle
\thispagestyle{empty}

\begin{abstract}
    We construct static, spherically symmetric, charged traversable wormhole solutions to the Einstein--Maxwell equations, supported by bidirectional (ingoing and outgoing) null dust with negative energy, and discuss a scenario for their dynamical formation from a black hole. 
    Our solution contains a traversable throat, where the areal radius takes a minimum, although the spacetime is not asymptotically flat.
    In our formation scenario, the spacetime evolves sequentially from a black hole to Vaidya regions and finally to a wormhole, with each transition mediated by an impulsive null shell.
    We find that the radius of the wormhole throat is determined by the mass and charge of the initial black hole as well as those of the injected shell.
\end{abstract}

\newpage
\setcounter{page}{1}

\tableofcontents

\section{Introduction}

The study of traversable wormholes has a long history in gravitational physics, originating as theoretical curiosities and evolving into a central topic in quantum gravity and spacetime topology.
Early theoretical models include the Einstein–Rosen bridge~\cite{einstein1935particle}, which connects two asymptotically flat spacetimes via a non-traversable geometry, and the Ellis–Bronnikov wormhole~\cite{ellis1973ether, ellis1979evolving, bronnikov1973scalar}, which is a traversable solution supported by phantom scalar fields. 
Since the seminal work of Morris and Thorne~\cite{Morris:1988cz, PhysRevLett.61.1446}, traversable wormholes have been investigated not only as potential tools for interstellar travel but also as important theoretical constructs illuminating the interplay between geometry, energy conditions, and quantum effects. Classical wormhole solutions within general relativity invariably require matter sources that violate the (averaged) null energy condition ((A)NEC)~\cite{Friedman:1993ty, Hochberg:1998ii, Graham:2007va}, such as exotic matter fields or negative-energy fluxes~\cite{Morris:1988cz, PhysRevLett.61.1446, Visser:1995cc, PhysRevD.56.4745}. 
This energy condition violation appears to be a generic feature of traversable wormholes, at least in the context of Einstein gravity.

Traditional wormhole solutions often assume some exotic matter which violates NEC at the classical level, while usual classical matter obeys it. 
One way to create traversable wormhole structure without assuming such exotic matter was introduced by Maldacena, Milekhin, and Popov~\cite{Maldacena:2018gjk}. 
They used a quantum mechanical force called Casimir effect as a source of negative energy. 
This effect appears when the wave function is confined to a finite space. 
To utilize it, they assumed the system consists of Einstein gravity, U(1) gauge field (magnetic field), and massless Dirac fermion coupled to gauge field in the usual way. 
Then they considered the setup of a traversable wormhole with two mouths placed apart in the asymptotically flat spacetime, just as Wheeler's picture~\cite{Wheeler:1955zz}. 
A magnetic field flows through the tunnel and outsides the mouths, which forms a loop pathway. 
The massless Dirac fermion under this background gauge fields gives a series of Landau levels, whose lowest level has zero energy. 
These states lead to two-dimensional chiral fermions in time and radial directions, trapped in the loop of magnetic field which has finite length. 
If one prepares somewhat strong magnetic fields, then the chiral fermions indicate a sufficiently large negative energy as a result of two-dimensional Casimir effect. 

Motivated by this discussion, we will study a traversable wormhole under the following setup. 
We assume that the system consists of Einstein gravity, U(1) gauge field, and bidirectional (ingoing and outgoing) negative energy null dust to support the wormhole structure. 
Here negative energy null dust is expected to play a similar role to Casimir energy as a source of wormhole structure near the throat: the energy-momentum tensor of two-dimensional Casimir energy has the form $T^{\mu \nu} \sim \mathrm{diag} (\epsilon, \epsilon, 0, 0) \quad (\mu, \nu = \{t, r, \theta, \varphi\})$ with negative $\epsilon$~\cite{Birrell:1982ix, Milton:1999ge}
\footnote{Of course this is very sketchy analysis. The concrete form of stress tensor depends on the metric. }
, which can be reproduced by 
bidirectional null dust: $T^{\mu \nu} = \epsilon (v^\mu v^\nu + u^\mu u^\nu)$, where $v^\mu$ and $u^\mu$ is bidirectional null vectors. 
Traversable wormhole supported by this negative energy dust was discussed in Ref.~\cite{Hayward:2002pm}. We will extend their work by including electromagnetic field.

Another main topic of this paper is to discuss the dynamical construction of the above wormhole geometry, starting from a black hole solution. This question is motivated by the idea that traversable wormholes supported by negative energy might arise from more familiar initial data, such as a Reissner–Nordström black hole. Understanding such a process can help to clarify the connection between black hole physics and semiclassical wormhole solutions.

Several works have studied dynamical wormhole formation~\cite{Koyama:2004uh, Hayward:2009yw, Maeda:2008bh}. 
Especially, Hayward showed that static wormholes can be supported by negative energy null dust~\cite{Hayward:2002pm}, and later, Koyama and Hayward constructed wormholes dynamically from Schwarzschild black holes using negative energy null shell and continuous null dust~\cite{Koyama:2004uh}. 
These works showed that violations of the energy condition can lead to traversable wormholes forming from black hole spacetimes.

In this paper, we extend these ideas to charged spacetimes under the above setup. 
We consider the dynamical formation of a charged wormhole by matching a black hole region, Vaidya-type regions, and a final wormhole region along null shells, following the method of Barrabès and Israel~\cite{Barrabes:1991ng}.

The plan of the remainder of this paper is as follows.
In Sec.~\ref{chawhs}, we construct
a solution of a charged traversable wormhole which is static and spherically symmetric, supported by negative energy null dust. 
Then, we will investigate the properties of the solution, such as throat structure, flare-out condition, and asymptotic geometry. 
In Sec.~\ref{neutfl}, we discuss a process of wormhole formation starting from two-sided Reissner-Nordstr\"{o}m (RN) black hole solution. 
This process can be realized by inserting an impulsive null shell, followed by a continuous stream of negative energy null dust that supports the wormhole throat. 
In that section we focus on the case where the initial black hole and final wormhole have the same electric charge. 
In Sec.~\ref{charfl}, we will generalize the discussion to the case where the electric charge becomes different along the spacetime evolution. 
This process can be realized when the null dust has some electric charges
\footnote{Note that in this case the electromagnetic duality is violated. }
, that is closer to the original Casimir effect setup in Ref.~\cite{Maldacena:2018gjk}. 
We summarize and discuss the results in Sec.~\ref{sumdis}. 

Throughout the paper, we use the geometrical unit in which the speed of light, the gravitational Newton constant, and the Coulomb constant are unity, that is, $c = G = k = 1$, where the Einstein--Hilbert--Maxwell action is expressed as
\begin{align}
S = \int d^4 x \sqrt{-g} \left[ \frac{1}{16 \pi} R - \frac{1}{16\pi} F_{\mu\nu} F^{\mu\nu}
\right] + S_{\mathrm{matter}}.
\end{align}

\section{The wormhole solution with electric field} \label{chawhs}

\subsection{The equations}

In this section, we construct a wormhole solution sourced by an electric field and a negative energy in the form of bidirectional null dust. 
We only consider the case where the spacetime is static and spherically symmetric. 
Thus, the metric and energy-momentum tensor have the following form by using arbitrary functions $f(r)$, $\psi(r)$ and $\epsilon(r)$
\footnote{We remark that our solution is not included in the series in Ref.~\cite{Kim:2024mam}, where they assumed asymptotically flatness. 
We will see that we cannot construct an asymptotically flat wormhole with this setup.}
;
\begin{align}
    ds^2 = - e^{2 \psi(r)} f(r) dt^2 + \frac{dr^2}{f(r)} + r^2 d \Omega^2, 
  \label{gean}
\end{align}
and
\begin{align}
  \tensor[]{T}{^\mu _\nu} =&~ \tensor[]{T}{_{\mathrm{(EM)}} ^\mu _\nu} + \tensor[]{T}{_{\mathrm{(dust)}} ^\mu _\nu},  \\
	\tensor[]{T}{_{\mathrm{(EM)}} ^\mu _\nu} =&~ \frac{Q^2}{8 \pi r^4} \mathrm{diag} (-1, -1, 1, 1), \\ 
  \tensor[]{T}{_{\mathrm{(dust)}} ^\mu _\nu} =&~ \mathrm{diag}\,(-\epsilon(r), \epsilon(r), 0, 0). 
\end{align}
Note that Eq.~\eqref{gean} is the most general form of a static and spherically symmetric metric. 
$\tensor[]{T}{_{\mathrm{(EM)}} ^\mu _\nu}$ is the energy-momentum tensor of electromagnetic fields, which has the form 
$ F^{tr} = e^{-\psi(r)} Q/r^2$
. 
It is the general form of the solution of Maxwell's equation in static and spherically symmetric spacetime, and $Q$ is a constant of integration which we will regard as electric charge (see App.~\ref{emtemf}). We assume $Q>0$ without loss of generality. 
$\tensor[]{T}{_{\mathrm{(dust)}} ^\mu _\nu}$ is the energy-momentum tensor of bidirectional null dust, which, we assume, has negative energy $\epsilon < 0$. 
Since $\tensor[]{T}{^\mu _\nu}$ satisfies the conservation law $\tensor[]{T}{^\mu ^\nu _{; \nu}} = 0$, 
the energy density of bidirectional null dust should have the form \cite{Hayward:2002pm}
\begin{equation}
  \epsilon(r) = - \frac{\lambda}{4 \pi r^2} 
  \frac{e^{-2\psi(r)}}{f(r)}
  , \label{endn}
\end{equation}
where $\lambda$ is a positive constant. 

By expressing the function $f(r)$ by another function $B(r)$ as  
\begin{align}
f(r) = 1 + \frac{Q^2}{r^2} - \frac{B(r)}{r},
\end{align}
the Einstein equation $G = 8 \pi T$ can be written as follows;
\begin{align}
 - \frac{B' (r)}{r^2} &= \frac{2 \lambda}{r^2} \frac{e^{-2\psi(r)}}{f(r)},  \label{defa}\\
  \frac{2 Q^2 - r B(r)}{r^4} + \left( 2 \psi'(r) + \frac{f'(r)}{f(r)} \right) \frac{f(r)}{r} &= - \frac{2 \lambda}{r^2} \frac{e^{-2 \psi(r)}}{f(r)} .
\end{align}
These equations come from the $\tensor[]{}{^t_t}$ and $\tensor[]{}{^r_r}$ components respectively. 
From Eq.~\eqref{defa}, the function $\psi(r)$ is determined as
\begin{equation}
  e^{- 2\psi(r)} = \frac{- B'(r)}{2 \lambda} \left( 1 + \frac{Q^2}{r^2} - \frac{B(r)}{r} \right). 
\end{equation}
Eliminating $\psi(r)$, we get 
\begin{equation}
  r (r^2 + Q^2 - r B(r)) B''(r) - (2 Q^2 - r B(r)) B'(r) + (r B'(r))^2 = 0. \label{deqb} 
\end{equation}
By multiplying $1 / r^3$, we can rewrite this as 
\begin{equation}
  \dv{r} \qty[ \qty(1 + \frac{Q^2}{r^2} - \frac{B(r)}{r}) B'(r) ] = -2 \frac{(B'(r))^2}{r} .\label{prpb}
\end{equation}
This equation is difficult to solve analytically, but in principle we can determine the solution except for the two integral constant degrees of freedom. 
Note that this solution also satisfies $\tensor[]{}{^\theta_\theta}$ and $\tensor[]{}{^\phi_\phi}$ components of Einstein equation.

Now, let us summarize the solution.
The metric has the form 
\begin{equation}
  ds^2 = \frac{2 \lambda}{B'(r)} dt^2 + \frac{1}{1 + \frac{Q^2}{r^2} - \frac{B(r)}{r}} dr^2 + r^2 d \Omega^2, \label{sssw}
\end{equation}
and the energy density of the bidirectional null dust is given by 
\begin{equation}
  \epsilon(r) = \frac{B'(r)}{8 \pi r^2}. 
\end{equation}
Since the time component of the metric must be negative, $B'(r)$, and hence $\epsilon(r)$, must also be negative. 
We will confirm numerically that a class of solutions satisfied the desired property in Sec.~\ref{exthra}. 
 
In order for the spacetime to be a ``wormhole", it should meet some conditions. 
First and foremost, it must have a throat, that is a two-sphere of radius $r_0$, where the geometry has a minimal areal radius. 
Since we use the areal radius itself as a radial coordinate, the metric must have a coordinate singularity there. 
Thus,
the radius of the throat is characterized by the condition $f(r_{0}) = 1 + Q^2 / r_0^2 - B(r_0) / r_0 = 0$. 
Then, the range of the radial coordinate is $r \in [r_0, \infty)$. 
The presence of the throat is confirmed by the numerical calculations in Sec.~\ref{exthra}, 
and here we assume that it is present.

By the definition of throat radius, the value of $B(r_0)$ at the throat can be expressed as 
 \begin{align}
      B(r_0) = r_0 + \frac{Q^2}{r_0}. \label{throat}
 \end{align}
Provided that $B''(r_{0})$ is finite, 
Eq.~\eqref{deqb} at $r=r_0$ is reduced to 
\begin{equation}
  \begin{gathered}
    (2 Q^2 - r_0 B(r_0)) B'(r_0) - (r_0 B'(r_0))^2 = 0.
  \end{gathered}
\end{equation}
Therefore, we gain the value of the derivative of $B(r)$ at the throat as 
\begin{align}
  B'(r_0) = \frac{Q^2}{r_0^2} - 1. 
\end{align}
Since $B'(r_{0})$ is negative, the radius of the throat must satisfy $r_{0} > Q$.

\subsection{The wormhole properties}

Next, to call the spacetime a traversable wormhole, the following conditions should be fulfilled~\cite{Morris:1988cz, PhysRevLett.61.1446}. 
\begin{align}
  &\text{Wormhole domain}:& & \text{The coordinate $r$ is well-defined for $r \in (r_{0}, \infty)$}.  \label{abho}\\
  &\text{Finite proper length} :& & \int_{r_0} \sqrt{g_{rr}} \dd{r} < \infty. \\
  &\text{Flare-out condition} :& &\left.\frac{\mathrm{d}^2 r}{\mathrm{d} r_{*}^{2}}\right|_{r = r_{0}} > 0. \label{foco}\\
  &\text{Regularity of throat} 
  :& & |R|, |R_{\mu \nu} R^{\mu \nu}|, |R_{\mu \nu \rho \sigma} R^{\mu \nu \rho \sigma}| < \infty \qquad \text{at} ~ r = r_0.   \\
  &\text{Asymptotically flatness} :& & g_{tt} \rightarrow -1, g_{rr} \rightarrow 1 \quad \text{as} ~r \rightarrow \infty. \label{asfl}
\end{align}
Here we defined the tortoise coordinate $r_* := \int \sqrt{- \frac{g_{rr}}{g_{tt}}} dr$ to describe the flare-out condition \eqref{foco}. 
The condition \eqref{abho} is confirmed by numerical calculation in Sec.~\ref{exthra}. 
Now we will check the remaining conditions. 

It will turn out that all the conditions are satisfied except for the condition \eqref{asfl}, i.e., it is not asymptotically flat, although our spacetime possesses the asymptotic region with diverging areal radius and the throat structure.
Therefore, in this sense, we will call this geometry a traversable wormhole.


Let us compute the proper distance around the throat in the constant time hypersurface
\begin{equation}
  \int_{r_0} \sqrt{g_{rr}} \dd{r} = \int_{r_0} \frac{1}{\sqrt{1+\frac{Q^2}{r^2} - \frac{B(r)}{r}}} \dd{r}. \label{prod2}
\end{equation}
Expanding $r$ near the throat
\begin{equation}
  r \simeq r_0 + Q \varepsilon,
\end{equation} 
with $\varepsilon \ll 1$ , we have
\begin{align}
  \frac{Q^2}{r^2} &\simeq \frac{Q^2}{r_0^2} \qty(1 - 2 \frac{Q}{r_0} \varepsilon), \\
  \frac{1}{r} &\simeq \frac{1}{r_0} \qty(1 - \frac{Q}{r_0} \varepsilon), \\
  B(r) &\simeq B(r_0) + B'(r_0) Q \varepsilon = \qty(r_0 + \frac{Q^2}{r_0}) + \qty(\frac{Q^2}{r_0^2} - 1) Q \varepsilon. 
\end{align} 
Therefore, Eq.~\eqref{prod2} is evaluated as 
\begin{equation}
  \begin{aligned}
    \int_{r_0} \sqrt{g_{rr}} \dd{r} &\simeq \int_{0} \frac{1}{\sqrt{- 2 \frac{Q}{r_0} \qty(\frac{Q^2}{r_0^2} - 1) \varepsilon}} \dd{\varepsilon}
     \propto \int_0 \frac{1}{\sqrt{\varepsilon}} \dd{\varepsilon}. 
  \end{aligned}
\end{equation}
This means that proper distance to the throat is finite. 
Note that this calculation ensures that the throat is located at a finite $r_{*}$, since $g_{tt}(r_{0})$ is assumed to be finite.

\subsubsection{Flare-out condition}

We can write down the flare-out condition using $r_{*}$;
\begin{equation}
  \left. \frac{\mathrm{d}^2 r}{\mathrm{d} r_{*}^{2}} \right|_{r = r_0} > 0. \label{flaout2}
\end{equation}
From the definition of $r_{*}$, we obtain 
\begin{align}
    \dv{r}{r_{*}} &= f(r) e^{\psi(r)} = \sqrt{\frac{2 \lambda (1 + Q^2 / r^2 - B / r)}{- B'(r)}}  \label{eq:drdrst}\\
    & \xrightarrow{(r \rightarrow r_0)} 0. \notag
\end{align}
Thus, the second derivative is
  \begin{align}
    \frac{\mathrm{d}^2 r}{\mathrm{d} r_{*
    }^{2}} 
    &= \dv{r}{r_*} \qty(\dv{r} \dv{r}{r_*}) \notag\\
    &= 2 \lambda \frac{r \left(r^2 + Q^2 -r B(r)\right) B''(r)+\left(2 Q^2-r B(r) +r^2 B'(r)\right) B'(r)}{2 r^3 B'(r)^2} \notag\\
    &= 2 \lambda \frac{2 Q^2-r B(r)}{r^3 B'(r)}  \label{eq:ddrdrst}\\
    & \rightarrow \frac{2 \lambda}{r_0} \quad (r \rightarrow r_0).\notag
  \end{align}
Since $r_0 > Q > 0$, we conclude that the flare-out condition \eqref{flaout2} is satisfied. 

As a prerequisite of this, the (averaged) null energy condition should be violated. 
We can also make sure it. 
Considering null vector for radial direction $\displaystyle k^\mu = \qty(1, \pm \sqrt{- \frac{g_{tt}}{g_{rr}}}, 0,0)$, the energy condition is
\begin{equation}
  \begin{aligned}
    \tensor[]{T}{_{\mathrm{(dust)}} ^\mu_\nu} k_{\mu} k^{\nu} &= (-\epsilon - \epsilon) g_{tt} \\
    &= - 2 \qty(- \frac{\lambda}{4 \pi r^2} a(r)) \qty(- \frac{1}{a(r)}) \\
    &= - \frac{\lambda}{2 \pi r}. 
  \end{aligned} \label{nec2}
\end{equation}
Here we used Eqs.~\eqref{gean} and \eqref{endn}. 
This is always negative, so the null energy condition is always violated.

\subsubsection{Curvatures at throat} 
We can check that the wormhole geometry has no singularity by calculating invariant curvature scalars;
\begin{align}
  R &= 0, \label{eq:R}\\
  R_{\mu \nu} R^{\mu \nu} &= \frac{2 r^4 B'(r)^2+4 Q^4}{r^8}, \label{eq:RR}\\
  C_{\mu\nu\rho\sigma} C^{\mu\nu\rho\sigma} &= \frac{12 (r B(r) - 2 Q^2)^2}{r^8}. \label{eq:CC} 
\end{align}
We used Eq.~\eqref{deqb} and its derivative to remove the second and third derivatives of $B(r)$. 
These are obviously not singular except for $r = 0$, that is outside of the domain. 
The curvatures at the throat $r = r_0$ are respectively, 
\begin{align}
  R |_{r = r_0} &= 0, \\
  R_{\mu \nu} R^{\mu \nu} |_{r = r_0} &= \frac{2\, \qty(3 Q^4 - 2 Q^2 r_0^2 + r_0^4)}{r_0^8}, \\
 C_{\mu\nu\rho\sigma} C^{\mu\nu\rho\sigma}|_{r = r_{0}} &= \frac{12 (r_{0}^2 - Q^2)^2}{r_{0}^8}. 
\end{align}

\subsubsection{Asymptotic behavior}

In the $r \rightarrow \infty$ limit, the $\displaystyle \frac{Q^2}{r^2}$ terms in metric \eqref{sssw} and the terms proportional to $Q^2$ in the differential equation \eqref{deqb} vanish.
This implies that the geometry approaches the wormhole without electric charge, which is discovered by Hayward \cite{Hayward:2002pm}, 
\begin{align}
  ds^2 &= - \frac{2 \lambda}{1 + 2 l \phi e^{l^2}} dt^2 + \frac{1 + 2 l \phi e^{l^2}}{2 \phi^2 e^{2 l^2}}dr^2  +r^2 d\Omega^2, \label{hayw} \\
  r &= a (e^{-l^2} + 2 l \phi), \label{defl} \\
  \phi(l) &:= \int_0^l e^{- l'^2} \dd{l'} - \frac{\sqrt{\pi}}{2} + \frac{\bar{r}}{2 a}. \label{phai}
\end{align}
Here, $l$ is a radial coordinate associated with $r$ through Eq.~\eqref{defl} and $a$ and $\bar{r}$ corresponds to two  integration constants, which are assumed to be positive.
The integral in $\phi$ is proportional to the error function.
In the asymptotic region, $l$ is proportional to $r$: $l \sim r / \bar{r}$.  

In the original setup by Hayward, which corresponds to $Q = 0$ in our case, $l$ is a global coordinate and it ranges $(- \infty, \infty)$.
In addition, when $\bar{r} = a \sqrt{\pi}$, it describes a symmetric wormhole 
and $l = 0$ corresponds to the throat with the areal radius $r_{0} = a = \bar{r}/\sqrt{\pi}$. 

Comparing the radial component of two metrics, we find that $B(r)$ should behave as
\begin{equation}
  B(r) \sim - \frac{1}{2 a} \bar{r}^2 e^{r^2/\bar{r}^2}. \label{hayb}
\end{equation}
The derivative of $B(r)$ can be evaluated as 
\begin{align}
    B'(r) \sim - \frac{r}{a} e^{r^2/\bar{r}^2}. \label{haybd}
\end{align}

Now let us investigate the asymptotic behavior of Eq.~\eqref{hayw}. 
In the $r \rightarrow \infty$ limit, or equivalently in the $|l|\rightarrow \infty$ 
limit, 
the metric clearly does not reduce to flat metric;
\begin{equation}
    ds^2 \sim - \frac{2 \lambda a}{r} e^{- r^2/\bar{r}^2} dt^2 + \frac{2 a r}{\bar{r}^2} e^{ - r^2/\bar{r}^2} dr^2 + r^2 d \Omega^2.
\end{equation}
Moreover, the infinite areal radius $r \to \infty$ is not infinitely far away. 
To see this, let us introduce again the $r_{*}$ coordinate, which can be evaluated as
\begin{align}
r_{*} \sim \frac{1}{\sqrt{\lambda}} \int^{r} \frac{r'}{\bar{r}} d r' = \frac{1}{2\sqrt{\lambda}} \frac{r^2}{\bar{r}}   + \text{const.} ~,
\end{align}
in the asymptotic region.
Then, outgoing null geodesic is expressed as a line of constant $t - r_{*}$.
Parameterizing this line by $r_{*}$, the tangent vector can be obtained as 
\begin{equation}
  k^\alpha = \qty( \frac{1}{\sqrt{\lambda}} \frac{r(r_{*})}{\bar{r}}, 1, 0, 0). 
\end{equation} 
This tangent vector satisfies non-affinely parameterized geodesic equation, 
$\tensor[]{k}{^\alpha _{;\beta}} k^\beta = \kappa(r) k^\alpha$ with $\kappa(r) = - 2 r/\bar{r}^2$.
Thus an affine parameter $\tau$ can be obtained as
\begin{equation}
  \begin{aligned}
    \dv{\tau}{r} = \exp(\int^{r} \kappa(r') \dd{r'}) = e^{-r^2/\bar{r}^2}, \\
    \therefore ~ \tau = \int^r e^{-r'^2/\bar{r}^2} \dd{r'} = \bar{r} \phi(r/\bar{r}) + \mathrm{const.}
  \end{aligned}
\end{equation}
This is finite in $r \rightarrow \pm \infty$ limit. 
This means that the affine parameter along the outgoing null geodesic from the throat to $r \to \infty$ is finite, which means that $r \to \infty$ is actually not null infinity. 

Finally we investigate the curvature scalars. 
By plugging the asymptotic expressions \eqref{hayb} and \eqref{haybd} into Eqs.~\eqref{eq:R}, \eqref{eq:RR}, and  \eqref{eq:CC}, we obtain 
\begin{align}
  R &= 0, \\
  R_{\mu \nu} R^{\mu \nu}
  & \sim
  \frac{2}{a^2 r^2} e^{2 r^2/\bar{r}^2} \to \infty,
  \label{risq} \\
  C_{\mu \nu \rho \sigma} C^{\mu \nu \rho \sigma} 
  &\sim \frac{3 \bar{r}^4}{a^2 r^6} e^{2 r^2/\bar{r}^2} \to \infty.
  \label{kret}
\end{align}
Both Eqs.~\eqref{risq} and \eqref{kret} diverge in $r \rightarrow \infty$ limit.
Thus, this spacetime is not only non-asymptotically flat, but also has curvature singularities in the asymptotic region~\footnote{Lack of asymptotically flatness was mentioned in Ref.~\cite{Hayward:2002pm}. However, they did not refer to the singularity.}.

\subsubsection{Global coordinates and Penrose diagram}
The above analysis suggests that one can introduce global coordinates by using $r_{*}$ instead of the areal radius $r$.
By choosing the origin of $r_{*}$ appropriately, we set $r_{*} = 0$ at the throat without loss of generality.

\begin{figure}[t]
  \centering
  \includegraphics[scale = 0.6]{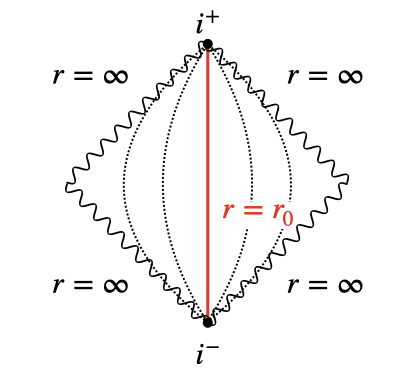}
  \caption{The Penrose diagram of the traversable wormhole solution.
  The wavy lines represent curvature singularities, where the areal radius diverges, $r \to \infty$. 
  This spacetime possesses future/past timelike infinities $i^{\pm}$ in the sense that there exist future/past complete timelike geodesics, such as the world line staying at the throat $r = r_{0}$ (shown as the red line).}
  \label{whpenr1}
\end{figure}

In the $r_{*}$ coordinates, the metric can be expressed as 
\begin{align}
ds^2 &= f(r) e^{2 \psi(r)} (- d t^2 + d r_{*}^2) + r(r_{*})^2 d \Omega^2 \\
&= - \frac{2 \lambda}{B'(r(r_{*}))} (- dt^2 + d r_{*}^2) + r(r_{*})^2 d \Omega^2.
\label{eq:ds2 global}
\end{align}
Although the original radial coordinate $r$ spans the region $r_{*} \in (0,\infty)$, the metric is now regular beyond $r_{*} = 0$.
Hence, now the metric \eqref{eq:ds2 global} is defined for $r_{*} \in (-\infty, \infty)$.
Since the $t,r_{*}$ part of the metric is conformally isometric to the two dimensional Minkowski metric $ - dt^2 + d r_{*}^2$, its Penrose diagram has the same shape as that of two dimensional Minkowski spacetime as shown in Fig.~\ref{whpenr1}.
As demonstrated in the previous subsections,
this spacetime is not asymptotically flat,  and $r_{*} \to \pm \infty$, where the areal radius is infinite, are curvature singularities. However, in contrast to a non-traversable wormhole such as the interior of a black hole, timelike observers can remain in this spacetime for an infinite proper time.
For example, the timelike curve at $r = r_{0}$ with fixed angular coordinates is a geodesic with proper time $t$, which extends to $t \to \infty$.
Thus, this spacetime can be interpreted as an eternal wormhole. Moreover, this is traversable because one can freely traverse the throat from any point of the spacetime.

\subsection{Numerical solutions}
\label{exthra}
One can rewrite the equations of motion for $B(r)$ in terms of $r(r_{*})$. By solving Eqs.~\eqref{eq:drdrst} and \eqref{eq:ddrdrst} for $B(r)$ and $B'(r)$, we obtain
\begin{align}
 B(r(r_{*})) &= \frac{1}{\sqrt{A(r_{*})}}\left( Q^2 + A(r_{*}) - \frac{(A - Q^2) A'(r_{*})^2}{2 A(r_{*}) A''(r_{*})} \right), \label{eq:BtoA}\\
 B'(r(r_{*})) &= - 4 \lambda \frac{A(r_{*}) - Q^2 }{A(r_{*}) A''(r_{*})} \label{eq:BdtoA}.
\end{align}
Here we introduce $A(r_{*}) := r(r_{*})^2$.
Then, by equating the $r$ derivative of Eq.~\eqref{eq:BtoA} with Eq.~\eqref{eq:BdtoA}, we obtain the differential equation for $A$ as  
\begin{align}
A' \left( \frac{A'''}{A''} + \frac{3 Q^2 - A}{2 (Q^2 - A)} \frac{A'}{A} \right) + 4 \lambda - A'' = 0.
\end{align}
This equation is symmetric under the constant shift and flip the sign of $r_{*}$, thus $r_{*} \to r_{*} + \text{const.}$ and $r_{*} \to - r_{*}$.
Naively, one might expect any wormhole solution to be symmetric because the presence of the throat, $A'(0) = 0$, appears to be a symmetric initial condition. However, since the coefficient of $A'''$ vanishes when $A' = 0$, the differential equation is structurally singular there. As a result, $A'''(0)$ is not determined from the data at $r_{*} = 0$, potentially breaking the symmetry. A symmetric solution can be obtained only when $A'''$ vanishes in the limit $r_{*} \to 0$,
\begin{align}
\lim_{r_{*} \to 0} A''' = \lim_{r_{*} \to 0} \frac{4 \lambda - A''} {A'} A'' = 0.
\end{align}
The numerical analysis shows that we obtain both symmetric and asymmetric wormhole solutions depending on the initial conditions. 

The parameter dependence in the differential equations can be eliminated by rescaling $A$ and $r_{*}$ as
\begin{align}
A &= Q^2 \hat{A},\\
r_{*} &= \frac{Q}{\sqrt{\lambda}} \hat{r}_{*}.
\end{align}
Then, the differential equation can be simplified as 
\begin{align}
\hat{A}' \left( \frac{\hat{A}'''}{\hat{A}''} + \frac{3 - \hat{A}}{2 (1 - \hat{A})} \frac{\hat{A}'}{\hat{A}} \right) + 4 - \hat{A}'' = 0.
\end{align}
From the analysis of asymptotic behavior in the previous subsection, we obtain  
\begin{align}
A \sim 2 \sqrt{\lambda} \bar{r} r_{*}, \quad  \Leftrightarrow
\quad 
\hat{A} \sim  2 \frac{\bar{r}}{Q} \hat{r}_{*}. \label{eq:Ahatasym}
\end{align}
Thus, the slope of $\hat{A}$ in the asymptotic region represents the scale $\bar{r}$ in the unit $Q$.

\begin{figure}[H]
\centering
  \includegraphics[width=0.8\textwidth]{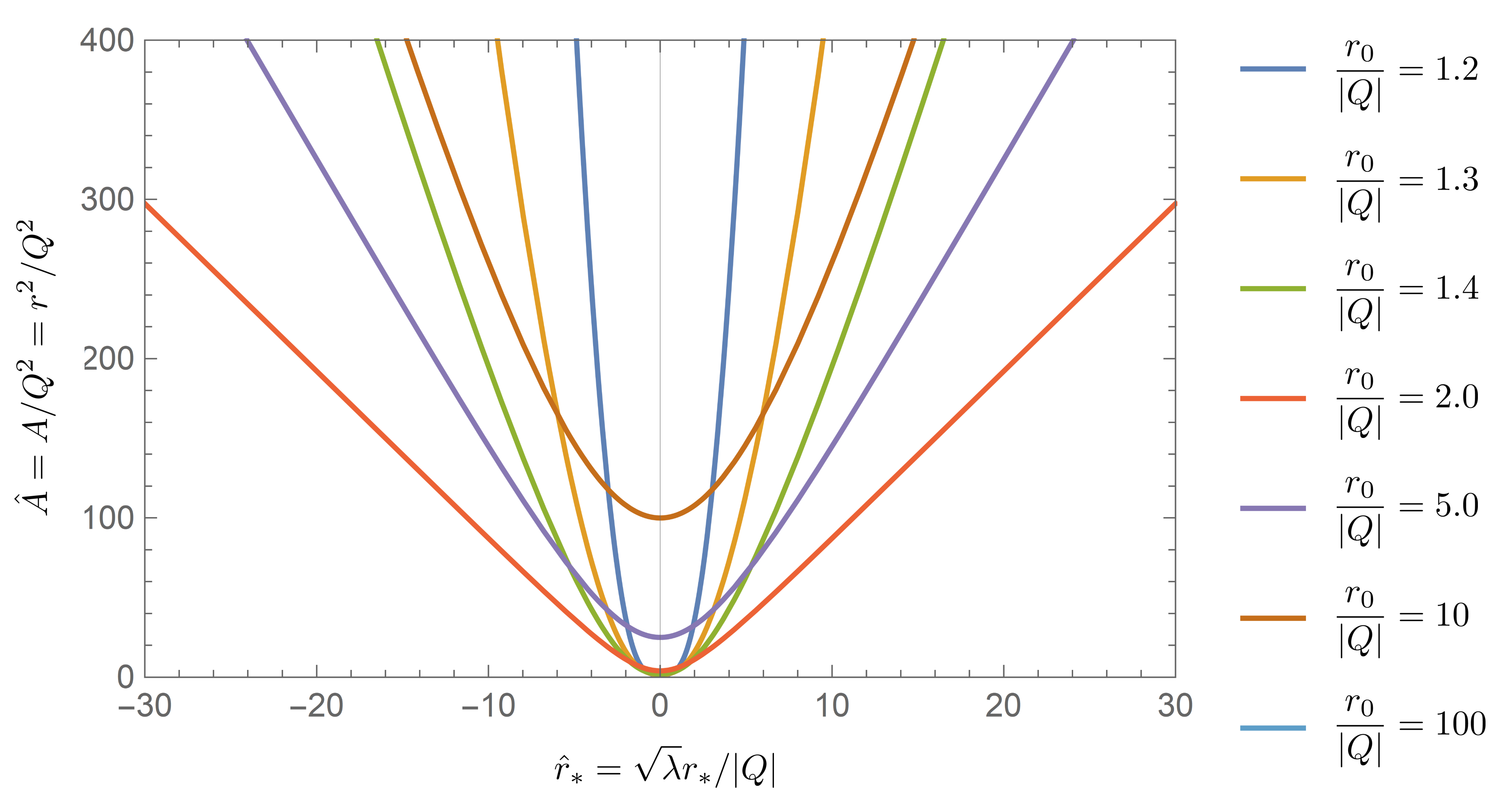}
  \caption{A series of symmetric wormhole solutions. 
  The initial conditions for each solution is set as $\hat{A}(0) = r_{0}/Q, \hat{A}'(0) = 10^{-5}, \hat{A}''(0) = 4$. Note that actual throat is located at $\hat{r}_{*} = \mathcal{O}(10^{-6})$.
  }
  \label{fig:Avsr}
  \end{figure}

\begin{figure}[H]
  \centering
  \includegraphics[width=0.8\textwidth]{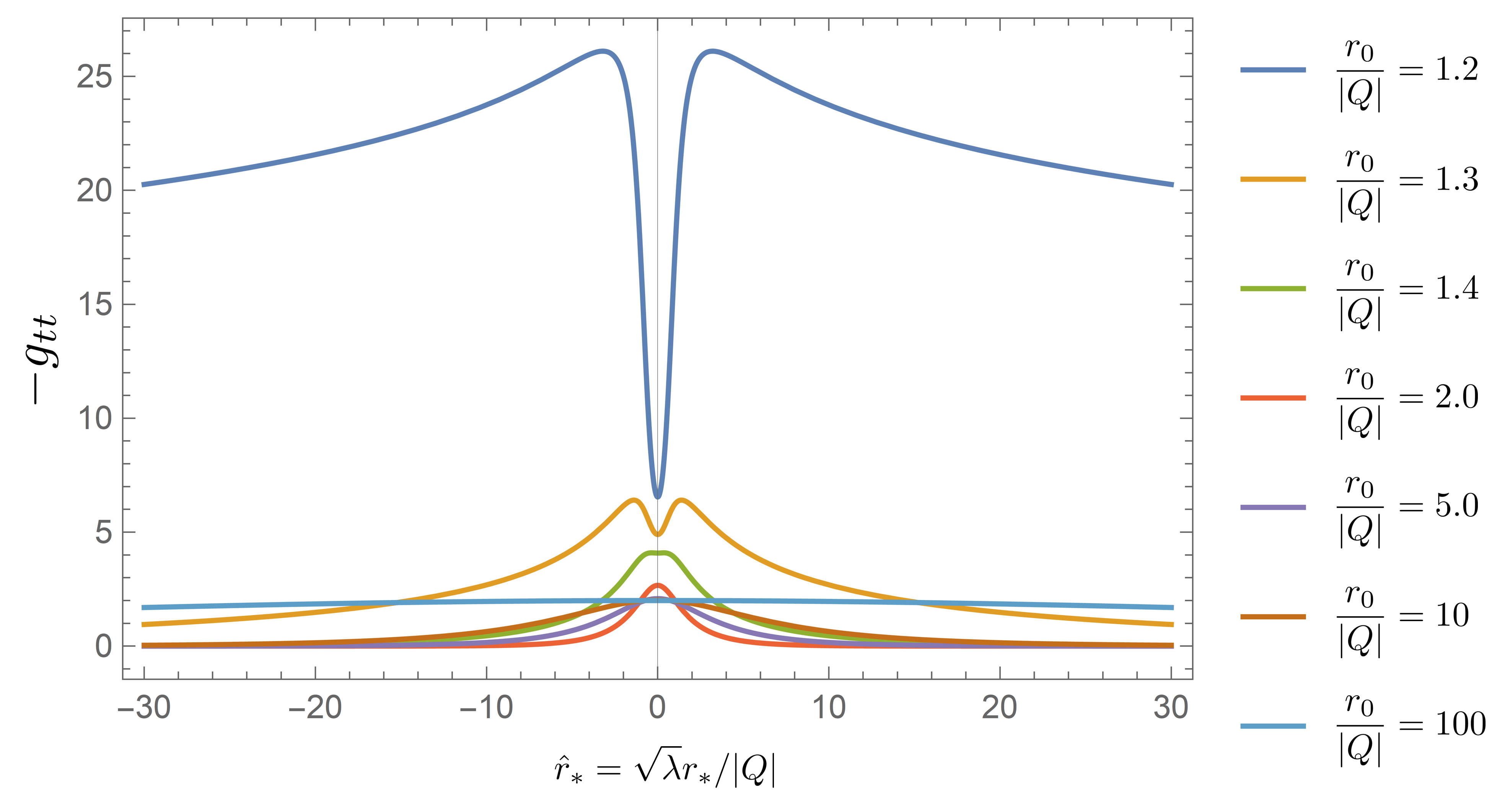}
  \caption{ Plot of $- g_{tt}$ for a series of symmetric wormhole solutions. All solutions satisfy $- g_{tt} > 0$ throughout the entire domain. }
  \label{fig:gttvsr}
\end{figure}

\begin{figure}[H]
  \centering
  \includegraphics[width=0.8\textwidth]{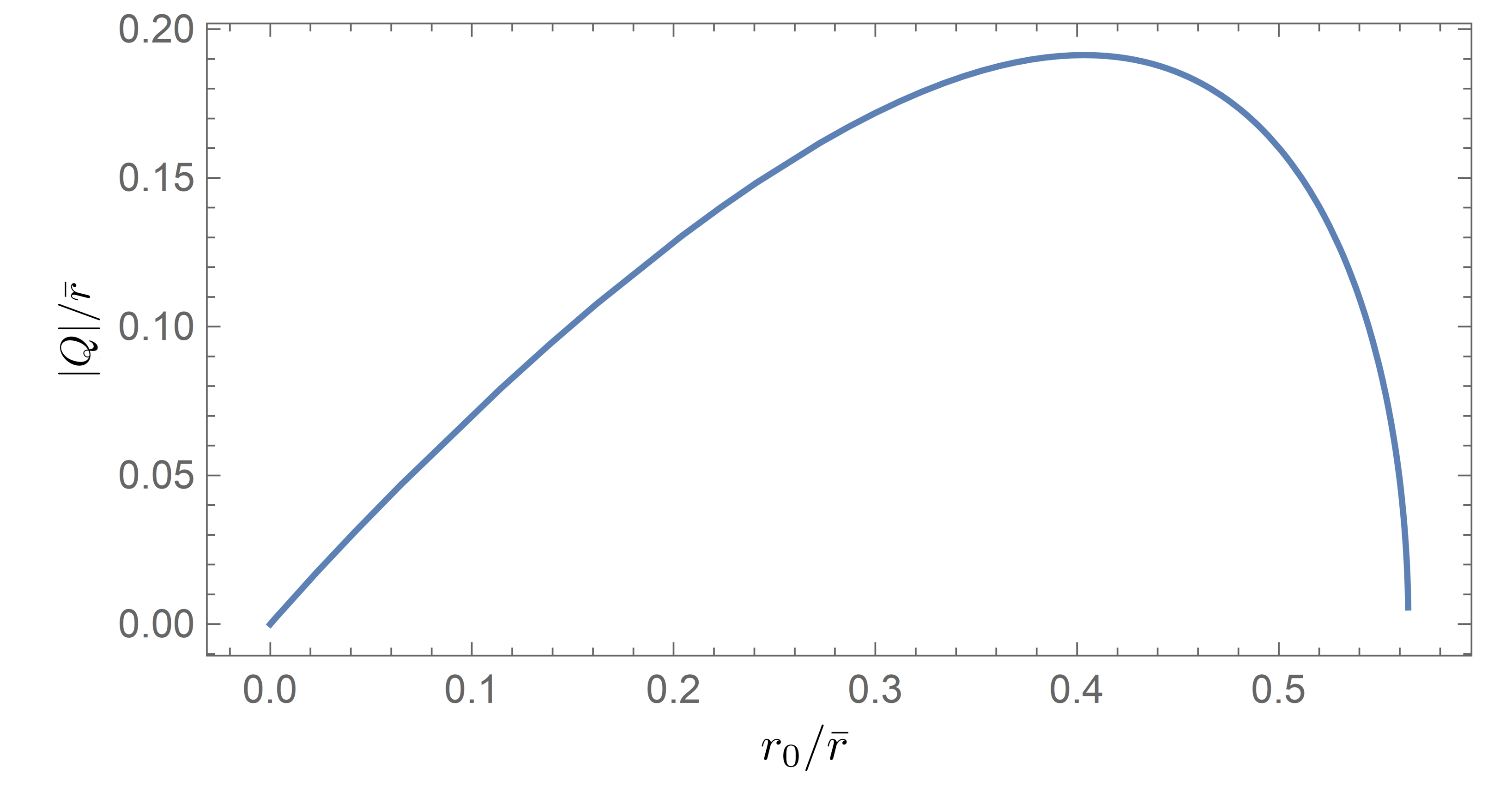}
  \caption{ Relation between $Q$ and $r_{0}$, fixing the asymptotic scale $\bar{r}$.
  The right edge of the plot approaches to $Q \to 0$ and $r_{0}/\bar{r} \to 0.564 \sim 1/\sqrt{\pi}$, which corresponds to the Hayward's symmetric wormhole solution. There is the upper bound for the charge $Q$, which is given by $Q/\bar{r} \sim 0.191$ with the throat size $r_{0}/\bar{r} \sim 0.403$. This corresponds to the initial condition $r_{0}/Q \sim 2.11$. 
  }
  \label{fig:Qvsr0}
\end{figure}

In Figs.~\ref{fig:Avsr} and ~\ref{fig:gttvsr}, we present a series of symmetric wormhole solutions varying an initial condition $r_{0}/Q > 1$. 
One can see that the solutions have desired properties: $- g_{tt}$, and hence $- B'(r)$, is positive for all solutions with $r_{0}/Q > 1$.
As expected, $\hat{A}$ becomes linear in the asymptotic region. For a given initial condition $r_{0}/Q$, the asymptotic slope of each plot allows us to determine the scale $\bar{r}/Q$ for each numerical solution. This, in turn, enable us to obtain the relation between $Q/\bar{r}$ and $r_{0}/\bar{r} = r_{0}/Q \times Q/ \bar{r}$ for each solution. The results of numerical calculations with initial conditions ranging from $r_{0}/Q = 1.2$ to $r_{0}/Q = 100$ are summarized in Fig.~\ref{fig:Qvsr0}. There, the asymptotic slope is extracted from the value at $\hat{r}_{*} = 10^8$.
One can find that, with fixing the asymptotic scale $\bar{r}$, the maximum size of the throat is achieved by $Q = 0$, which corresponds to Hayward's symmetric solution, Eq.~\eqref{hayw} with $\bar{r} = a \sqrt{\pi}$, where the throat size is given by $r_{0}/\bar{r} = 1/ \sqrt{\pi}$ analytically. From the plot, we find that there is an upper bound for the charge, which is given by $Q/\bar{r} \sim 0.191$, which corresponds to the initial condition $r_{0}/Q \sim 2.11$. We can not obtain symmetric wormhole solution with the charge larger than this value. This behavior is similar to Reissner-Nordstr\"{o}m (RN) black hole, where it has event horizons at small charge, but there is no horizon when it has a sufficiently large charge. 
The maximum value of the charge is expected to correspond to the extremal value in the black hole case. 

\begin{figure}
\centering
\includegraphics[width=\textwidth]{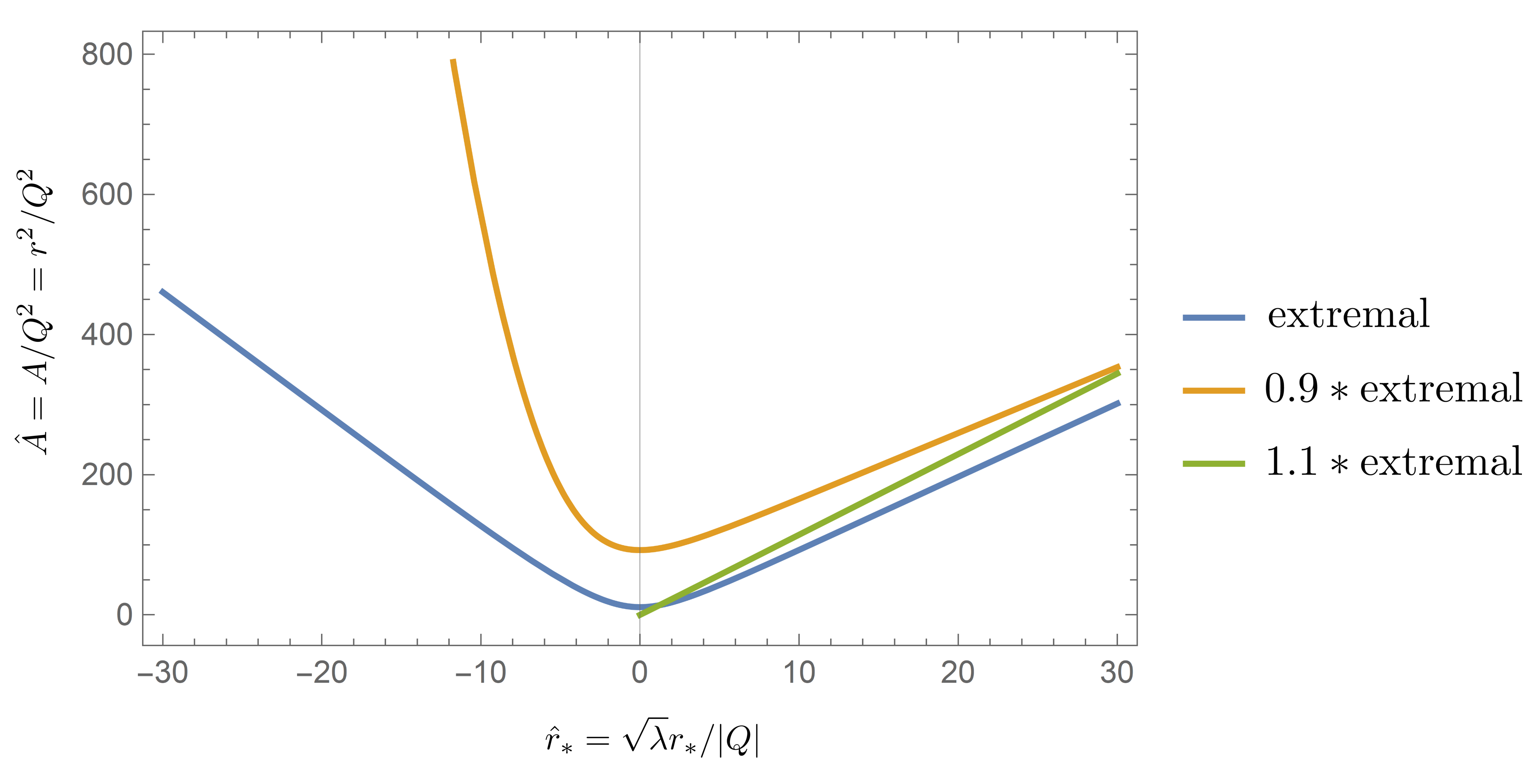}
  \caption{
  Solutions with the following initial conditions imposed at $\hat{r} = 50$:
  (i) $(\hat{A}, \hat{A}', \hat{A}'')= (\hat{A}_{\text{ext}}, \hat{A}'_{\text{ext}}, \hat{A}''_{\text{ext}})$ (blue),
  (ii)
   $(\hat{A}, \hat{A}', \hat{A}'') = (\hat{A}_{\text{ext}}, 0.9 *\hat{A}'_{\text{ext}}, \hat{A}''_{\text{ext}})$ (orange),
   (iii)
   $(\hat{A}, \hat{A}', \hat{A}'') = (\hat{A}_{\text{ext}}, 1.1 *\hat{A}'_{\text{ext}}, \hat{A}''_{\text{ext}})$ (green), where $A_{\text{ext}}$ is the symmetric wormhole solution with the extremal initial condition $r_{0}/Q \sim 2.11$.
   The origin of $\hat{r}_{*}$ is shifted so that $\hat{r}_{*} = 0$ corresponding $A' = 0$. The orange plot represents asymmetric wormhole, while the green plot cannot be solved beyond $r_{*} = 0$.
  }
  \label{fig:Avsr_asym}
  \end{figure}

\begin{figure}
\centering
\includegraphics[width=\textwidth]{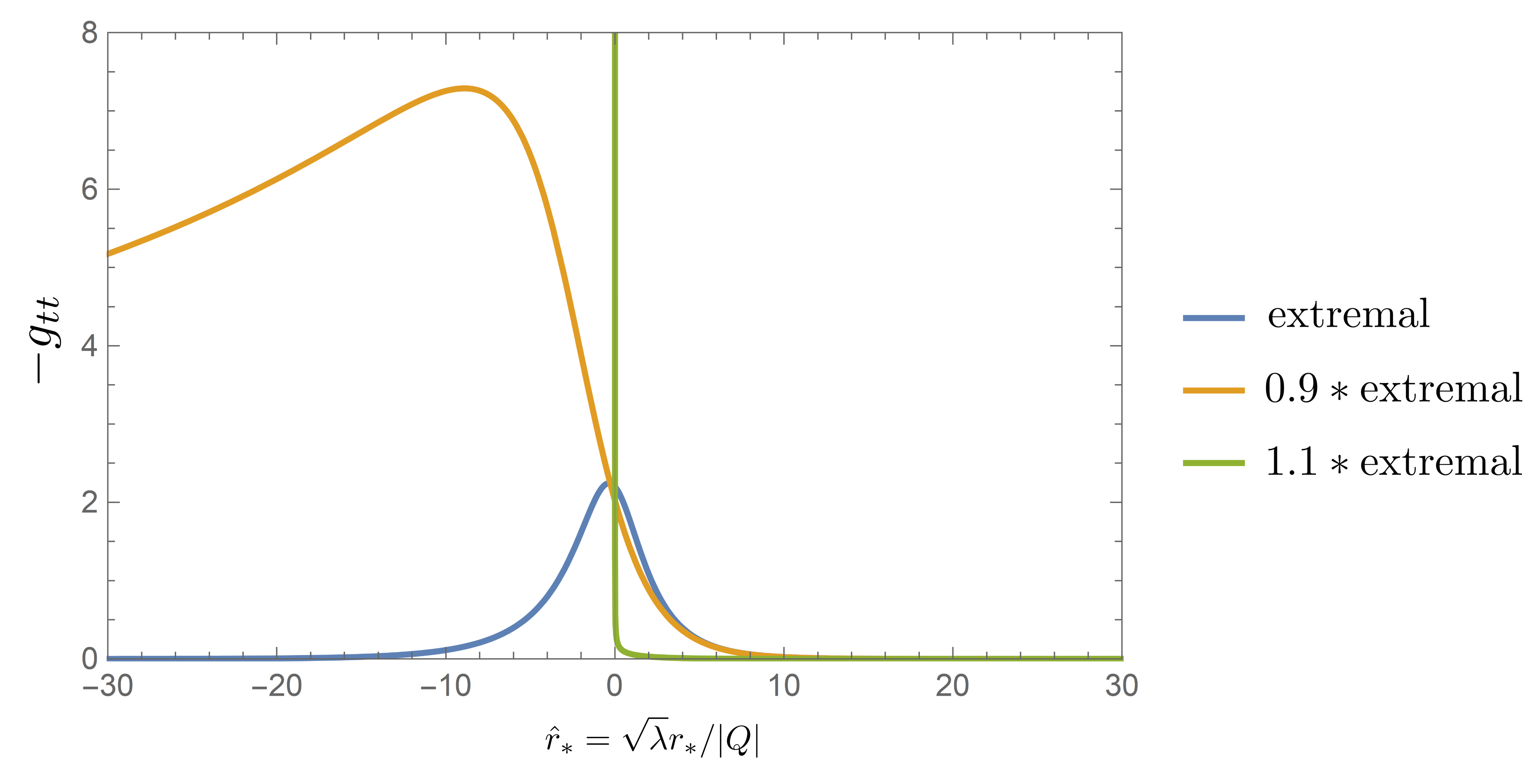}
  \caption{
  Plots of $-g_{tt}$ with the same initial conditions as Fig.~\ref{fig:Avsr_asym}.
  $g_{tt}$ is regular and has correct sign for wormhole solutions (blue and orange), but is singular in the other solution (green) in the $r_{*} \to 0$ limit. }
  \label{fig:mgttvsr_asym}
  \end{figure}

Finally, let us investigate solutions other than symmetric wormhole. In Figs.~\ref{fig:Avsr_asym} and \ref{fig:mgttvsr_asym}, we express an asymmetric wormhole solution and a singular solution. By reading the slope of each solution, we can again find the ratio of the charge $Q$ to the asymptotic scale $\bar{r}$, the value of which are $Q/\bar{r} \sim 0.213$ for the asymmetric wormhole solution and $Q/\bar{r} \sim 0.174$ for the singular solution.
We note that the value for the asymmetric wormhole is greater than our bound. Thus, the bound $Q/\bar{r} \leq 0.191$ should be understood as that only for symmetric wormholes. 

\section{Wormhole formation - neutral flux} \label{neutfl}

In the following of this paper, we shall
discuss formation of a wormhole starting from a black hole. 
First, we review 
three geometries: RN black hole, Vaidya spacetime, and traversable wormhole. 
Then we summarize the junction conditions for gluing these three spacetimes. 
Finally, we will discuss the construction procedure. 
In this section, we focus on the case where all three spacetimes have the same electric charge, and no electric current exists. 
We will discuss the case with charged current in the next section. 

\subsection{Geometries}

\subsubsection{Reissner-Nordstr\"{o}m black hole}

In the presence of only electromagnetic field, the static and spherically symmetric solution of Einstein equation is known as RN black hole: 
\begin{align}
  ds^2 &= - f_B(r) dt^2 + \frac{dr^2}{f_B(r)} + r^2 d \Omega^2, \\
  f_B(r) &= 1 - \frac{2M}{r} + \frac{Q^2}{r^2}. 
\end{align}
Here, $M$ represents the mass and $Q$ is the electric (or equivalently magnetic) charge located at the origin. 
The electric field is described by
\begin{equation}
  \tensor[]{F}{^t^r} = \frac{Q}{r^2}, 
\end{equation}
which satisfies Maxwell's equation. 
The energy-momentum tensor has the form
\begin{equation}
  \begin{aligned}
    \tensor[]{T}{^\mu_\nu} &= \frac{1}{4 \pi} \qty(F^{\mu \alpha} F_{\nu \alpha} - \frac{1}{4} \tensor[]{\delta}{^\mu_\nu} F^{\alpha \beta} F_{\alpha \beta}) \\
    &= \frac{Q^2}{8 \pi r^4} \mathrm{diag} (-1, -1, 1, 1) . \label{elemag}
  \end{aligned}
\end{equation}
Note that when $Q < M$,
there are two horizons
$r_{\pm}=M \pm \sqrt{M^2-Q^2}$.
$r_{+}$ and $r_{-}$ are called the outer and inner horizons, respectively. 
When $Q = M$, the locations of two horizons coincide, and the geometry is called extremal, beyond which there is no horizon, and a naked singularity appears at $r=0$. 
Later, we will see that our construction is valid only when $Q < M$, so we will focus on this case. 

For later calculations, it is convenient to take the Eddington-Finkelstein coordinates. 
The metric has the following form in this coordinate
\footnote{We can consider not only ingoing Eddington-Finkelstein coordinate but also outgoing ones by just flipping the sign before $2 dr$. This coordinate is also useful when one discuss other situations, for example, wormhole collapse into black hole. More detailed discussion can be found in Ref.~\cite{Koyama:2004uh} where they discuss wormhole construction and wormhole enlargement without electric fields.}
,
\begin{equation}
  \begin{aligned}
    ds^2 &= - e^{\psi_B} dv \qty(f_B e^{\psi_B} dv - 2 dr) + r^2 d \Omega^2, \\
    f_B &= 1 - \frac{2M}{r} + \frac{Q^2}{r^2}, \\
    \psi_B &= 0. 
  \end{aligned}
\label{efc}
\end{equation}
Here $v$ is the ingoing null coordinate defined by $v := t + r_*$, and $r_* := \int \dd{r} 1 / f(r)$ is the tortoise coordinate. 

\begin{figure}[H]
  \centering
  \includegraphics[scale = 0.5]{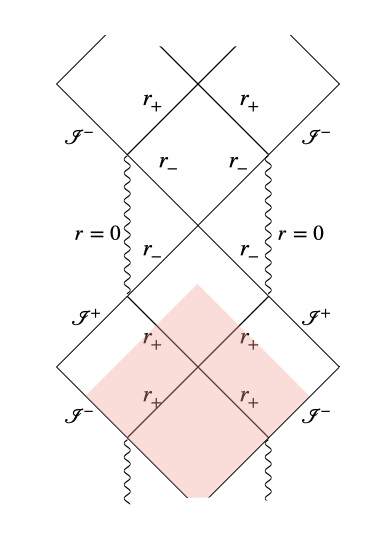}
  \caption{The Penrose diagram of a RN black hole. We will use the red region as the starting point of wormhole construction. }
  \label{rnpenr}
\end{figure}

The Penrose diagram of this spacetime is as in Fig.~\ref{rnpenr}. 

\subsubsection{Vaidya spacetime}

Next we review the Vaidya spacetime with constant electric charge;
\begin{equation}
  ds^2 = - \qty(1 - \frac{2 m(V)}{r} + \frac{Q^2}{r^2}) dV^2 + 2 dV \, dr + r^2 d\Omega^2 . \label{vaidya}
\end{equation}
This is a solution of the Einstein--Maxwell equation with neutral null dust, where the electric field is given by
\begin{align}
    F^{Vr} = \frac{Q}{r^2}.
\end{align}
The energy-momentum tensor is given by
\begin{align}
  T_{\mu \nu} = \tensor[]{T}{^{\mathrm{(EM)}} _{\mu \nu}} + \frac{1}{4 \pi r^2} \dv{m(V)}{V} \delta^V_{\mu} \delta^V_{\nu}. 
\end{align}
We assumed that only the mass $m$ would be changed by the null dust. 
The energy-momentum tensor of the electromagnetic field $\tensor[]{T}{^{\mathrm{(EM)}} _{\mu \nu}}$ has the same matrix form as Eq.~\eqref{elemag} in this case (see App.~\ref{emtemf} for the derivation). 
The second term of the energy-momentum tensor represents the single directed (ingoing in this case) 
negative energy null dust ($m'(V) < 0$). 

For later convenience, we rewrite the line element in the same form as in Eq.~\eqref{efc};
\begin{equation}
  \begin{aligned}
    ds^2 &= - e^{\psi_V} dV \qty(f_V e^{\psi_V} dV - 2 dr) + r^2 d \Omega^2, \\
    f_V &= 1 - \frac{2M(V)}{r} + \frac{Q^2}{r^2}, \\
    \psi_V &= 0. 
  \end{aligned}
\label{vaiefc}
\end{equation}

\subsubsection{Traversable wormhole}

The geometry of the traversable wormhole with electric charge $Q$ is what we discussed in the previous section. 

\begin{figure}[H]
  \centering
  \includegraphics[scale = 0.6]{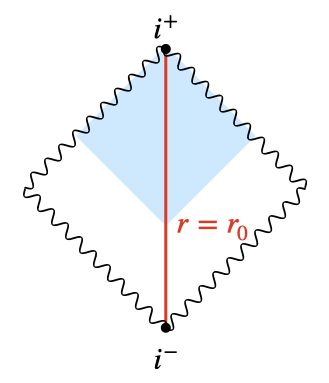}
  \caption{The Penrose diagram of the traversable wormhole solution.
  We will construct the blue region starting from the black hole geometry. }
  \label{whpenr2}
\end{figure}

Similarly as to the last two spacetimes, we rewrite the metric in Eddington-Finkelstein coordinate. 
\begin{align}
  ds^2 &= - e^{\psi_W} du \qty(f_W e^{\psi_W} du + 2 dr) + r^2 d \Omega^2, \label{whefc}\\
  f_W &= 1 + \frac{Q^2}{r^2} - \frac{B(r)}{r}, \\
  e^{\psi_W} &= \sqrt{- \frac{2 \lambda}{B'(r) (1 + Q^2 / r^2 - B / r)}}.
\end{align}
We remark that we took the outgoing null coordinate $u$ instead of the ingoing one $v$. 
The sign before $2 dr$ is flipped from Eq.~\eqref{efc}. 
Moreover, by defining double null coordinate as $v, u = t \pm r_*$, we can express the metric by
\begin{equation}
  ds^2 = \frac{2 \lambda}{B'(r)} du dv + r^2 d\Omega^2. 
\end{equation}
In the following subsection, we will consider a dynamical formation scenario constructing a spacetime region $u > u_{0}, v > v_{0}$, expressed as the blue region in Fig.~\ref{whpenr2}.

\subsection{The strategy}

\begin{figure}[H]
  \centering
  \includegraphics[scale = 0.5]{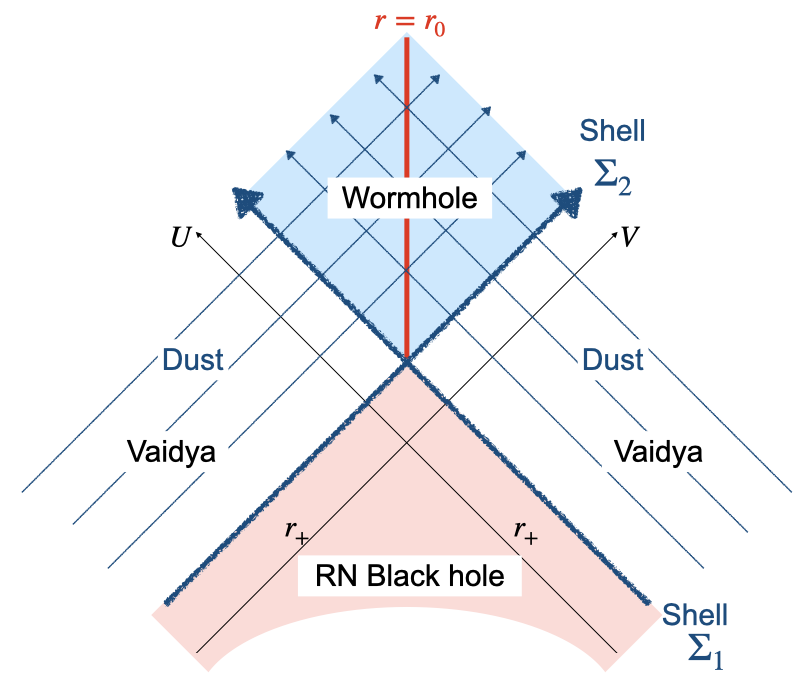}
  \caption{The schematic picture of wormhole formation. 
  We start from a black hole, and go to a wormhole via Vaidya regions. 
  The regions are separated by the null shells. }
  \label{whpenr}
\end{figure}

Now we will show our strategy 
for constructing  the wormhole.
The schematic picture of the wormhole formation is shown in Fig.~\ref{whpenr}. 
The spacetime begins from a two-sided (eternal) black hole, goes through Vaidya spacetime, and ends with a wormhole. 
A null dust shell ($\Sigma_1$ and $\Sigma_2$) is inserted between each geometry. 
The lower shell on the right side ($\Sigma_1$) is located upper (later time) than the past horizon in the black hole geometry, so that we have a usual asymptotically flat region before the wormhole formation. 
Note that this construction is the same as the previous work \cite{Koyama:2004uh} except for the presence of electric fields. 

The location of $\Sigma_1$ is identified by the scalar equation $v = v_0$ (const.) on the black hole side, and $V = V_0$ (const.) in the Vaidya region. 
On the other hand, the location of $\Sigma_2$ is represented as $u = u_0$ (const.) in the wormhole, and $V = V_1(r)$ surface (outgoing null coordinate becomes constant at this surface) in the Vaidya spacetime. 
The function $V_1(r)$ can be obtained from the Vaidya metric \eqref{vaiefc} as 
\begin{align}
  &f_V e^{\psi_V} dV - 2 dr = 0, \label{defv1}\\
  \Leftrightarrow \quad &\dv{V}{r} = \frac{2}{1 - \frac{2 m(V)}{r} + \frac{Q^2}{r^2}}.
\end{align}
Then, $V_1(r)$ is a solution of this equation. 

\subsection{Junction conditions} \label{juncon}

The next task is to consider the junction condition at these null shells. 
We will follow the discussion given by Barrabès and Israel in Ref.~\cite{Barrabes:1991ng}. 

We first summarize the notation and review the junction condition developed in Refs.~\cite{Barrabes:1991ng, Poisson:2009pwt}. 
Let us imagine matching two different geometries at null hypersurfaces $\Sigma$, and define the three dimensional coordinate $\{\xi^a\} ~ (a = 1,2,3)$ on the surface. 
The vielbein on the surface is defined by $e^\mu_{(a)} := \pdv{x^\mu}{\xi^a}$. 
We promise that $\xi^1$ is taken to be an arbitrary parameter of null generators of $\Sigma$. 
Then, $e^\mu_{(1)}$ is orthogonal to all $e^\mu_{(a)}$, including itself, which means it is also a normal vector to the hypersurface. 
From now on, we will denote $e^\mu_{(1)} =: n^\mu$, and other two vielbeins as $e^\mu_{(A)} ~ (A = 2,3)$. 
We define another vector called transverse null vector $N^\mu$ which satisfies $N^\mu N_\mu = e_{(A)}^\mu N_\mu = 0$ and $n^\mu N_\mu = -1$. 
Moreover, we define a scalar function $\Phi (x) $ that characterizes the hypersurface by $\Sigma: \Phi(x) = 0$. 
This enables us to represent the normal in another way:
\begin{equation}
  n_\mu = - \alpha^{-1} \partial_\mu \Phi, 
\end{equation}
where $\alpha$ is a normalization factor that confirms $- \alpha^{-1} \partial_\mu \Phi = g_{\mu \nu} e^\nu_{(a = 1)}$. 
We can determine it by $\alpha = N^\mu \partial_\mu \Phi$ .

Under this setup, we define the transverse extrinsic curvature:
\begin{equation}
  \mathcal{R}_{ab} := - N_\mu e^\nu_{b} \qty(\nabla_{\nu} e^\mu_{a}).
\end{equation}
Then, we should require the following junction conditions. 
\begin{enumerate}
  \item Continuity of induced metric 
  \begin{equation}
    [h_{AB}] := [g_{\mu \nu}] e_A^\mu e_B^\nu = 0 . \label{firjc}
  \end{equation}
  Here we defined the two-dimensional induced metric on the surface ${h}_{AB} := g_{\mu \nu} e^\mu_A e^\nu_B$. 
  In addition, we denote the difference of some quantity $X$ between upper side and lower side of hypersurface by $[X] := X|_{\Phi \rightarrow +0} - X|_{\Phi \rightarrow -0}$. 

  \item The surface stress tensor is identified to the discontinuity of the transverse curvature
  \begin{align}
    T^{\mu \nu}_{\mathrm{surface}} &= \alpha S^{\mu \nu} \delta (\Phi), \\
    S^{\mu \nu} &= \sigma n^\mu n^\nu + j^A \qty(n^\mu e^\nu_A + e^\mu_A n^\nu) + P h^{AB} e^\mu_A e^\nu_B, \\
    \sigma &= - \frac{1}{8 \pi} h^{AB} \qty[\mathcal{R}_{AB}], \\
    j^A &= \frac{1}{8 \pi} h^{AB} \qty[\mathcal{R}_{1 B}], \\
    P &= - \frac{1}{8 \pi} \qty[\mathcal{R}_{1 1}].
\end{align}
\end{enumerate}
From the first junction condition, we should identify the same areal radii points of the two spacetimes. 
We will now focus on the second junction condition
\footnote{We should also consider the junction of electric fields (See App.~\ref{elmgjc} for the detail). One can check that if all regions have the same electric charge, the condition is automatically satisfied.
We will deal with this problem in detail in the next section, including the case where each geometry has different charges. }
.

\subsubsection{Useful formulae}

We will review useful notations introduced by Koyama and Hayward \cite{Koyama:2004uh} 
\footnote{We take $\pm \zeta r$ as a surface parameter, while in Ref.~\cite{Koyama:2004uh} they choose $r$. Consequently, the definition of normal vector is changed to $n_\mu = - \alpha \partial_\mu \Phi$, which was defined as $n_\mu = \zeta \alpha \partial_\mu \Phi$ in original paper. After these changes, we will reach the correct final result \eqref{exv}. }
. 

All three spacetimes can be described in the following form (Eddington-Finkelstein coordinates) as we saw in Eqs.~\eqref{efc}, \eqref{vaiefc}, and \eqref{whefc};
\begin{equation}
  ds^2 = - e^{\psi} dv \qty(f e^{\psi} dv + 2 \zeta dr) + r^2 d \Omega^2. \label{genefc}
\end{equation}
In this subsection only, we shall uniformly use $v$ to denote null coordinates. $\zeta$ is a sign factor, which takes $-1$ when $v$ is an ingoing null coordinate and $+1$ when $v$ is an outgoing null coordinate. 

First, we consider the null hypersurface $\Phi =v - v_0 = 0$, where $v_0$ is some constant. 
This hypersurface corresponds to the null shells in three cases: 
(i) $\Sigma_1$ in black hole side, (ii) $\Sigma_1$ in Vaidya side, and (iii) $\Sigma_2$ in wormhole side. 
We choose the intrinsic coordinates to be $\xi^a = (\zeta r, \theta, \phi)$. 
The vielbein is
\begin{equation}
  e^\mu_{(a)} := \pdv{x^\mu}{\xi^a}= \{\zeta \delta^\mu_r, \delta^\mu_\theta, \delta^\mu_\varphi\}.
\end{equation}
Note that $e^\mu_{(1)} = n^\mu$. 
The normal and transverse null vectors are
\footnote{
Upper index form is
\begin{equation}
    N^\mu = e^{-\psi} \delta^\mu_v - \zeta \frac{f}{2} \delta^\mu_r.
\end{equation}
}
\begin{align}
  n_\mu &= - e^{\psi} \partial_\mu \Phi = - e^{\psi} \delta^v_\mu, \label{normal1}\\
  N_\mu &= - \frac{f e^{\psi}}{2} \delta^v_\mu - \zeta \delta^r_\mu. 
\end{align}
The normalization factor is determined to be $\alpha = e^{- \psi}$. 
The transverse extrinsic curvature is
\begin{equation}
    \mathcal{R}_{ab} = \mathrm{diag} \qty(\zeta \pdv{\psi}{r}, -\zeta \frac{rf}{2}, -\zeta \frac{rf}{2} \sin^2 \theta) . \label{exv} 
\end{equation}

Next we consider the other null hypersurface $\tilde{\Phi} = v - v_1(r) = 0$
, where $v_1 (r)$ is defined in the same way as Eq.~\eqref{defv1}. 
\begin{equation}
  \dv{v_1(r)}{r} = - \zeta \frac{2}{f} e^{-\psi}. 
\end{equation}
This hypersurface corresponds to the null shell $\Sigma_2$ in the Vaidya side. 
The argument of $f$ is $f(v=v_1(r), r)$. 
We choose the intrinsic coordinates $\xi^a = (- \zeta r, \theta, \phi)$. 
The vielbein is
\begin{equation}
  \tilde{e}^\mu_{(1)} = \frac{2}{f} e^{-\psi} \delta^\mu_v - \zeta \delta^\mu_r = \tilde{n}^\mu, \quad \tilde{e}^\mu_{(2)} = \delta^\mu_\theta, \quad \tilde{e}^\mu_{(3)} = \delta^\mu_\varphi. 
\end{equation}
The normal and transverse null vectors are respectively
\footnote{
Upper index form is
\begin{equation}
    \tilde{N}^\mu = \zeta \frac{f}{2} \delta^\mu_r.
\end{equation}
}
\begin{align}
  \tilde{n}_\mu &= - e^{\psi} \partial_\mu \tilde{\Phi} = - e^\psi \delta^v_\mu - \zeta \frac{2}{f} \delta^r_\mu , \label{norma2}\\
  \tilde{N}_{\mu} &= - \frac{f}{2} e^{\psi} \delta^v_\mu. 
\end{align}
Again the normalization is determined as $\alpha = e^{-\psi}$. 
The transverse extrinsic curvature is
\begin{equation}
  \mathcal{R}_{ab} = \mathrm{diag} \qty(- \frac{2 e^{-\psi}}{f^2} \pdv{f}{v} - \zeta \pdv{\psi}{r}, \zeta \frac{rf}{2}, \zeta \frac{rf}{2} \sin^2 \theta). \label{exu}
\end{equation}

\subsection{Matching geometries}

Let us move on to the discussion about wormhole construction. 

\subsubsection{Matching black hole and Vaidya}

Now we match the RN black hole and Vaidya region located in the right side in Fig.~\ref{whpenr}, across $\Sigma_1$. The junction to the left side Vaidya region can be performed similarly.
Using Eqs.~\eqref{efc}, \eqref{vaiefc}, and \eqref{exv}
(and choosing the sign to be $\zeta = -1$ in both sides)
, we get the surface energy
\begin{align}
  \sigma_1 &= \left. - \frac{f_V - f_B}{8 \pi r} \right|_{\Sigma_1} = \frac{m(V_0) - M}{4 \pi r^2}, \label{enden1}\\
  j^A_1 &= P_1 = 0. 
\end{align}
Here and in the following, the subscripts $1$ and $2$ of quantities of the surface energy momentum tensor represent those on the shell $\Sigma_1$ and $\Sigma_2$, respectively.
Thus, this shell actually consists of null dust (in other words, pressureless). 

The surface energy momentum tensor becomes 
\begin{equation}
  \begin{aligned}
    T_1^{\mu \nu} &= \alpha \sigma_1 n^\mu n^\nu \delta(V - V_0) \\
    &= \frac{m(V_0) - M}{4 \pi r^2} n^\mu n^\nu \delta(V - V_0) . \label{emten1}
  \end{aligned}
\end{equation}

\subsubsection{Matching Vaidya and wormhole}

Next we match the right Vaidya region and wormhole region on $\Sigma_2$.
Using Eqs.~\eqref{whefc}, \eqref{exv} for wormhole side (with $\zeta = +1$) and Eqs.~\eqref{vaiefc}, \eqref{exu} for the Vaidya side (with $\zeta = -1$), we obtain the surface energy and pressure. 
\begin{align}
  \sigma_2 &= \left. \frac{f_W - f_V}{8 \pi r} \right|_{\Sigma_2} \notag \\ 
  &= \frac{2 m(r) - B(r)}{8 \pi r^2}, \label{eneden}
\end{align}
\begin{align}
  P_2 &= \left. - \frac{1}{8 \pi} \qty[\pdv{\psi_W}{r} - \qty(-\frac{2 e^{-\psi_V}}{f_V^2} \pdv{f_V}{V}) ] \right|_{\Sigma_2} \notag \\
  &= - \frac{1}{8 \pi} \qty[\qty(-\frac{1}{2} \frac{\qty{B'(r) (1 + Q^2 / r^2 - B / r)}'}{B'(r) (1 + Q^2 / r^2 - B / r)}) - \frac{2}{r f_V} \dv{m(r)}{r}] \notag \\
  &= - \frac{1}{8 \pi} \qty[\qty(-\frac{1}{2} \frac{-2 (B'(r))^2 / r}{B'(r) (1 + Q^2 / r^2 - B / r)}) - \frac{2 / r}{(1 - 2 m / r + Q^2 / r^2)} \dv{m(r)}{r}] \notag \\
  &= \frac{1}{8 \pi r} \frac{1}{(1 - 2 m/r + Q^2 / r) (1 + Q^2 / r^2 - B / r)} \notag \\
  & \hspace{80pt} \times \qty[2 \qty(1 + \frac{Q^2}{r^2} - \frac{B(r)}{r}) \dv{m(r)}{r} - \qty(1 + \frac{Q^2}{r^2} - \frac{2 m(r)}{r}) \dv{B(r)}{r}] .
\end{align}
We defined $m(r) := m(V)|_{V = V_1(r)}$. From the second to third line of $P_2$, we used the differential equation \eqref{prpb}.

Since we are considering the null dust case, we require $P_2=0$;
\begin{equation}
  2 \qty(1 + \frac{Q^2}{r^2} - \frac{B(r)}{r}) \dv{m(r)}{r} - \qty(1 + \frac{Q^2}{r^2} - \frac{2 m(r)}{r}) \dv{B(r)}{r} = 0 . \label{pres}
\end{equation}
This equation determines the form of the mass function $m(r)$ on the connecting surface. 
Solving this, we get
\begin{equation}
  m(r) = \frac{1}{2} B(r) + C \sqrt{- \frac{B' (1 + Q^2 / r^2 - B / r)}{2 \lambda}}, \label{mass2}
\end{equation}
where $C$ is an integral constant. 
Since the mass function depends only on $V$, we can identify the energy flux in the whole Vaidya region, by solving $V = V_1(r)$ for $r$ , denoting $r = r_{1}(V) = V_{1}^{-1}(V)$. Substituting this into Eq.~\eqref{mass2} , we obtain  
\begin{equation}
  m(V) = \frac{1}{2} B(r_{1}(V)) + C \sqrt{-\frac{B'(r_{1}(V)) (1 + Q^2 / r_{1}(V)^2 - B(r_{1}(V)) / r_{1}(V))}{2 \lambda}}. \label{vaimas}
\end{equation}

Substituting Eq.~\eqref{mass2} into \eqref{eneden}, we get
\begin{equation}
  \sigma_2 = \frac{C}{4 \pi r^2} \sqrt{-\frac{B' (1 + Q^2 / r^2 - B / r)}{2 \lambda}}
  = \frac{C}{4 \pi r^2} e^{- \psi_{W}}
  . 
\end{equation}
Thus, we can express the surface energy-momentum tensor,
\begin{equation}
  \begin{aligned}
    T_2^{\mu \nu} &= e^{-\psi_W} \sigma_2 n^\mu n^\nu \delta(u - u_0) \\
    &= \frac{C}{4 \pi r^2} \qty(-\frac{B' (1 + Q^2 / r^2 - B / r)}{2 \lambda}) n^\mu n^\nu \delta(u - u_0) \\
    &= \sigma_2 n^\mu n^\nu \delta(V - V_1(r)) \\
    &= \frac{C}{4 \pi r^2} \sqrt{-\frac{B' (1 + Q^2 / r^2 - B / r)}{2 \lambda}}  n^\mu n^\nu \delta(V - V_1(r)) . \label{ssvw}
  \end{aligned}
\end{equation}

\subsection{The whole picture of wormhole formation}

In this subsection, we shall describe the whole picture of the spacetime. 

\begin{figure}[h]
  \centering
  \includegraphics[scale = 0.5]{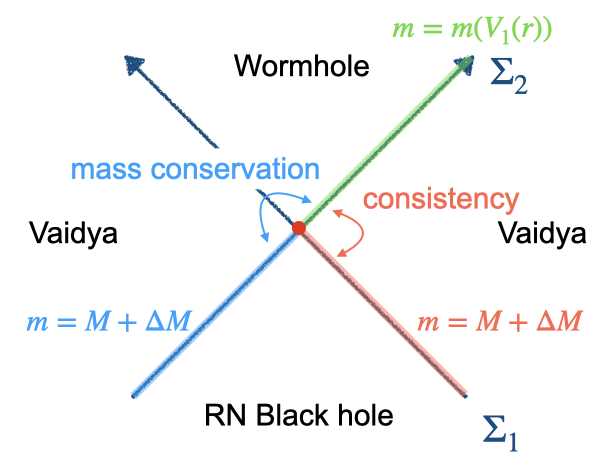}
  \caption{The way to determine the throat radius as a function of mass $M$, shell energy $\Delta M$, and charge $Q$. The identification of $m(V_0)$
  with $m(V_{1}(r_{0}))$ is obtained from green and orange shells, which gives Eq.~\eqref{thrrad}. Moreover, the mass conservation between blue and green shells at intersection point produces Eq.~\eqref{eq:continuity of T}, then we can identify the mass function of Vaidya region from the initial black hole spacetime.
  }
  \label{detrad}
\end{figure}

We start from the eternal RN black hole, and inject the null dust shells of mass $\Delta M$, corresponding to
energy density $\sigma_1 = \Delta M / 4 \pi r^2$. 
Immediately,
we turn on the negative energy null dust, then the geometry is connected to Vaidya spacetime. 
From Eq.~\eqref{enden1}, the mass function of the Vaidya region on the shell $\Sigma_1$ is
\begin{equation}
  m(V_0) = M + \Delta M . \label{shemas}
\end{equation}
On the other hand, the full behavior of the null dust in the Vaidya region is determined in Eq.~\eqref{vaimas}.
Since $V_{0}$ corresponds to the value of $V$ at the throat $r = r_{0}$ on $\Sigma_{2}$,
\begin{equation}
  m(V_0) = \frac{1}{2} B(r_0) = \frac{1}{2} \qty(r_0 + \frac{Q^2}{r_0}). 
\end{equation}
We identify this mass to Eq.~\eqref{shemas}, 
because both of them represent the mass function along the null shell $\Sigma_1$ 
(see Fig.~\ref{detrad}). 
From this condition, the throat radius is determined as
\begin{equation}
  \begin{aligned}
    & \frac{1}{2} \qty(r_0 + \frac{Q^2}{r_0}) = M + \Delta M, \\
    \Leftrightarrow~ & r_0 = (M + \Delta M) \pm \sqrt{(M + \Delta M )^2 - Q^2}. \label{thrrad}
  \end{aligned}
\end{equation}
Since the throat radius must be larger than the charge $r_0 > Q$,  we should choose the $+$ sign here. 
For the throat radius to be located between the inner and the outer horizon, the shell energy $\Delta M$ must be negative, 
which is compatible with our expectation that the injection of a negative energy is necessary for the formation of a wormhole from a black hole.
Moreover, $\Delta M$ must be larger than $Q - M ( < 0)$ so that the throat radius becomes real. 
This implies that
we cannot have an overcharged wormhole geometry: if we violate condition $\Delta M > Q - M$, we will have a naked singularity at $r=0$, just as in the case of RN black holes. 
We remark that this inequality is again consistent to the previous work \cite{Koyama:2004uh} with $Q \rightarrow 0$ limit. 

We can relate the throat radius and the timing of the null shell. 
Since we consider symmetric geometry, the throat is located at a $t = \text{const.}$ surface inside the outer horizon. 
Thus, the timing of the null shell $v = v_0$ 
is related to the throat radius $r_0$ via
\begin{equation}
  \begin{aligned}
    V_0 &= r_* (r = r_0) \\
    &= - r_0 + \frac{2 M^2 - Q^2}{2 \sqrt{M^2 - Q^2}} \log \qty(\frac{\sqrt{M^2 - Q^2} + (-M + r_0)}{\sqrt{M^2 - Q^2} - (-M + r_0)}) \\
    & \hspace{80pt} - M \log \left(\frac{-Q^2 + 2 M r_0 - r_0^2}{Q^2} \right),
  \end{aligned}
\end{equation}
up to constant shift. Here, $r_*$ is the tortoise coordinate of the RN black hole. To get the second line, we used a concrete form of the tortoise coordinate in the $r \in [r_-, r_+]$ region. 

Finally, we comment on the integral constant $C$, 
which is related to the energy density of null shell $\Sigma_2$. 
Since the left-going and right-going shells have no interaction, we can assume that their mass should be conserved respectively.
Such junctions beyond a crossing point is studied, for example, in Ref.~\cite{Nakao:1999qto} for timelike shells. 
For null shells, we require continuity of the surface energy-momentum tensor across the crossing point.
We provide a detailed analysis of continuity in App.~\ref{sec:app C}. 
There, we find that continuity of the energy-momentum tensor requires
 $\Delta M = - (-B'(r_{0})/\lambda) C^2/r_{0} < 0$, as shown in Eq.~\eqref{eq:continuity of T} in the appendix.

Let us summarize the result. 
We constructed a traversable wormhole from a black hole via Vaidya spacetime. 
The throat size of the created wormhole is determined by the three parameters on the initial black hole and injected shell. 
Furthermore, $\Delta M$ must be in the range $Q - M < \Delta M < 0$.

\section{Wormhole formation - charged flux} \label{charfl}

In this section, we will consider wormhole formation by using charged dust as a pulse. 
Along with that, the Vaidya region includes charged flux. 
We discuss the possibility of formation of a charged wormhole from a black hole that has less (or more) charge, and from a Schwarzschild black hole as the zero charge limit. 
We do not assume $Q > 0$ in this section, and allow the wormhole and black hole to have different sign charges.
Note that we assume that there is no charged flux in the wormhole region. 
We will leave the wormhole with the charged flux case for future work. 

\subsection{Junction conditions of electromagnetic field}

At the beginning, we review the junction condition of electromagnetic field in curved spacetime. 
This is the generalization of the usual junction condition discussed in classical electromagnetism, where the matching surface is chosen to be timelike. 
Here we will consider the null hypersurface case. 
Setup is the same as in Sec.~\ref{juncon}; on the hypersurface $\Phi = 0$, we define normal vector $n_\mu = - \alpha^{-1} \partial_\mu \Phi$, other tangent vectors $e^\mu_{(A)}$, and transverse vector $N^\mu$. 
The junction condition is
\begin{equation}
  \qty[F_{\mu \nu}] n^{\mu} = i_\nu, \label{nullco}
\end{equation}
where $i^\mu$ is related to the surface current:
\begin{equation}
  j^\mu = - \alpha ~i^\mu ~\delta(\Phi) . 
\end{equation}
We will give the details of the derivation in App.~\ref{elmgjc}. 
The condition Eq.~\eqref{nullco} can also be written in coordinate invariant form: 
\begin{align}
  \qty[F_{\mu \nu}] n^{\mu} e^\nu_{(A)} &= i_\nu e^\nu_{(A)}, \\
  \qty[F_{\mu \nu}] n^{\mu} N^\nu &= i_\nu N^\nu . 
\end{align}
Note that $i_\nu n^\nu$ is trivially zero.

\subsection{Charged Vaidya spacetime}

The charged Vaidya spacetime is written in Eddington-Finkelstein coordinates as Refs.~\cite{Bonnor:1970zz, Koh:2020hta}
\begin{equation}
  \begin{aligned}
    ds^2 = - f(v,r) dV^2  + 2 dV dr + r^2 d \Omega^2 , \\
    f(v,r) = 1 - \frac{2 m(V)}{r} + \frac{q(V)^2}{r^2}. 
  \end{aligned}
\end{equation}
Electromagnetic potential and field strength are
\begin{align}
  A &= \qty(\frac{q(V)}{r} - \frac{q(V)}{R}) \dd{v}, \\
  F &= \frac{q(V)}{r^2} \dd{v} \wedge \dd{r}, 
\end{align}
where $R$ is a gauge freedom and we can choose any value. The energy-momentum tensor and the electric current, which satisfy the Einstein equation and Maxwell equation $\nabla^\mu F_{\mu \nu} = j_\nu$ are 
\begin{align}
  \tensor[]{T}{^\mu ^\nu} =&~ \tensor[]{T}{_{\mathrm{(EM)}} ^\mu ^\nu} + \tensor[]{T}{_{\mathrm{(dust)}} ^\mu ^\nu}, \\
  \tensor[]{T}{_{\mathrm{(EM)}} ^\mu _\nu} =&~ \frac{q(V)^2}{8 \pi r^4} \mathrm{diag}\,(-1, -1, 1, 1), \\
  \tensor[]{T}{^{\mathrm{(dust)}} _V _V} =&~ \frac{2}{8 \pi r^3} (r m'(V) - q(V) q'(V)), \\
  j_V =& - \frac{q'(V)}{r^2}. 
\end{align}

\subsection{Wormhole formation with charged shell}
Now we discuss the formation of wormhole with charged dust. 
The general strategy is the same as the the previous section, but we can vary the electric charge of the black hole (we will denote $Q_{\mathrm{ini}}$) from that of the wormhole (we will denote $Q_{\mathrm{fin}}$). 
Furthermore, we will use the charged Vaidya spacetime in the middle stage of construction. 
The metric in Eddington-Finkelstein coordinates is respectively
\begin{align}
  &\text{Black hole} & : &
  \setlength{\abovedisplayskip}{100pt}{
    \begin{cases}
    ds^2 = - e^{\psi_B} dv (f_B e^{\psi_B} dv - 2 dr) + r^2 d \Omega^2, \\
    f_B = 1 - \frac{2M}{r} + \frac{Q^2_{\mathrm{ini}}}{r^2}, \qquad \psi_B = 0, \label{chbhf}
  \end{cases}
  } \\
  \vspace{10pt}
  &\text{Vaidya} & : & 
  \begin{cases}
    ds^2 = - e^{\psi_V} dV (f_V e^{\psi_V} dV - 2 dr) + r^2 d \Omega^2, \\
    f_V = 1 - \frac{2 m(V)}{r} + \frac{q(V)^2}{r^2} , \qquad \psi_V = 0, \label{chvaf}
  \end{cases} \\
  &\text{Wormhole} & : &
  \begin{cases}
    ds^2 = - e^{\psi_W} du (f_W e^{\psi_W} du + 2 dr) + r^2 d \Omega^2, \\
    f_W = 1 + \frac{Q_{\mathrm{fin}}^2}{r^2} - \frac{B(r)}{r}, \\
    e^{\psi_W} = \sqrt{-\frac{2 \lambda}{B'(r) (1 + Q^2_{\mathrm{fin}} / r^2 - B / r)}}. \label{chwof}
  \end{cases} 
\end{align}
The electromagnetic field strength and charged current in these coordinates are respectively
\begin{align}
  &\text{Black hole} & : & F^{v r} = \frac{Q_{\mathrm{ini}}}{r^2}, \qquad j^\mu = 0, \label{chbhem} \\
  &\text{Vaidya} & : & F^{v r} = \frac{q(V)}{r^2}, \qquad j_v = - \frac{\dot{q} (v)}{r^2}, \\
  &\text{Wormhole} & : & F^{v r} = e^{- \psi_W} \frac{Q_{\mathrm{fin}}}{r^2}, 
  \qquad j^\mu = 0. \label{chwoem} 
\end{align}

We will first discuss the matching between the black hole and the charged Vaidya regions, then between the Vaidya and the wormhole regions. 
Finally, we analyze the whole geometry. 

\subsubsection{Junction between black hole and Vaidya region}
The metric junction condition gives the surface energy and pressure as
\begin{align}
  \sigma_{1} &= - \frac{[f]}{8 \pi r} = 
  \frac{1}{8 \pi r} \left( \frac{2m(V_0) - 2M}{r} - \frac{q(V_0)^2 - Q_{\mathrm{ini}}^2}{r^2} \right), \label{sigma1} \\
  P_{1} &= 0 . 
\end{align}
This is obviously null dust. 
The junction of the electromagnetic fields provides the surface electric current
\footnote{The calculation is as follows. 
In Eddington-Finkelstein coordinate, we have
\begin{align}
  F_{\mu \nu} n^\mu N^\nu &= g_{\mu v} g_{\nu r} F^{v r} n^\mu N^\nu \notag\\
  &= (g_{rv})^2 F^{vr} (n^r N^v - n^v N^r). \label{useele}
\end{align}
Thus, 
\begin{equation}
  i_\nu N^\nu = [F_{\mu \nu} n^\mu N^\nu] = - \frac{q(V_0) - Q_{\mathrm{ini}}}{r^2}.
\end{equation}
Combining this with $i_\nu n^\nu = 0$, we get the answer. }
, 
\begin{equation}
  \begin{aligned}
      i_{1 \, \mu} &= \frac{q(V_0) - Q_{\mathrm{ini}}}{r^2} \delta^V_\mu \\
      &= - \frac{q(V_0) - Q_{\mathrm{ini}}}{r^2} n_\mu. 
  \end{aligned}
\end{equation}
This is a quite intuitive result: Gauss' law.

\subsubsection{Junction between Vaidya region and wormhole}
We repeat the same argument as in the previous section. 
The surface energy and pressure are given by
\begin{align}
  \sigma_{2} &= \frac{f_W - f_V}{8 \pi r} \notag \\ 
  &= \frac{1}{8 \pi r} \qty(\frac{2 m(r) - B(r)}{r} + \frac{Q_{\mathrm{fin}}^2 - q^2(r)}{r^2}),
\end{align}
\begin{align}
  P_{2} &= - \frac{1}{8 \pi} \qty[\pdv{\psi_W}{r} - \qty(-\frac{2 e^{-\psi_V}}{f_V^2} \pdv{f_V}{V}) ] \notag \\
  &= \frac{1}{8 \pi r} \frac{1}{(1 - 2 m/r + q^2 / r) (1 + Q_{\mathrm{fin}}^2 / r^2 - B / r)} \notag \\
  & \hspace{5pt} \times \qty[2 \qty(1 + \frac{Q_{\mathrm{fin}}^2}{r^2} - \frac{B(r)}{r}) \qty{\dv{m(r)}{r} - \frac{q(r)}{r} \dv{q(r)}{r}}- \qty(1 + \frac{q(r)^2}{r^2} - \frac{2 m(r)}{r}) \dv{B(r)}{r}] . 
\end{align}
Here we defined $q(r) := q(V = V_1(r))$. 
We demand that the pressure must vanish, which relates the mass function $m(r)$ and charge function $q(r)$ through the differential equation.
\begin{align}
m(r) = \frac{1}{2} B(r) + \frac{q(r)^2 - Q_{\text{fin}}^2}{2 r} + C(r) \sqrt{- \frac{B'(r) (1 + Q^2 / r^2 - B(r) / r)}{2 \lambda}}, \label{eq:m(r)charged}
\end{align}
where $C(r)$ is obtained as
\begin{align}
C(r) = C(r_{0}) + \int^{r}_{r_{0}} \frac{q(r')^2 - Q_{\text{fin}}^2}{ 2 r'^2 
\sqrt{ - \frac{B'(r')}{2 \lambda} \left(1 + \frac{Q_{\text{fin}}^2}{r'^2} - \frac{B(r')}{r'} \right) }} dr'.
\end{align}
The surface current is 
\footnote{In this case, from Eq.~\eqref{useele} we get
\begin{align}
  [F_{\mu \nu} n^\mu N^\nu] = \frac{Q_{\mathrm{fin}} - q(V_1(r))}{r^2}.
\end{align}}
\begin{equation}
  \begin{aligned}
      i_{2 \, \mu} &= \frac{Q_{\mathrm{fin}} - q(r)}{r^2} e^{- \psi_W} \delta^u_\mu \\
      &= - \frac{Q_{\mathrm{fin}} - q(r)}{r^2} n_\mu. 
  \end{aligned}
\end{equation}
Again, the current accounts for the discontinuity in the electric fields. 

\subsubsection{The whole geometry}
Let us move on to the analysis of the whole geometry. 
The entire scenario is the same as the previous setup and the only difference is the inclusion of the functional degree of freedom $q(V)$. The junction conditions leave $q(V)$ undetermined, allowing for a wide variety of formation scenarios depending on the choice of $q(V)$.

As in the previous setup, by setting $r = r_{0}$ in Eq.~\eqref{eq:m(r)charged}, we obtain 
\begin{align}
m(r_{0}) = \frac{1}{2} B(r_{0}) + \frac{q(r_{0})^2 - Q_{\text{fin}}^2}{2 r_{0}} = \frac{1}{2}\left( r_{0} + \frac{q(r_{0})^2}{r_{0}} \right),
\end{align}
from the junction $\Sigma_{2}$ between Vaidya and wormhole region.
Let us express the increase of the mass and charge due to the junction $\Sigma_{1}$ between black hole and Vaidya region, measured at $r \to \infty$, as $\Delta M$ and $\Delta Q$ respectively. Thus, we define 
\begin{align}
\Delta M &:= \lim_{r \to \infty} 4 \pi r^2 \sigma_{1}(r) = m(V_{0}) - M, \\
\Delta Q &:= q(V_{0}) - Q_{\text{ini}}.  
\end{align}
Since the limit $r \to r_{0}$ along $\Sigma_{2}$ approaches to $\Sigma_{1}$, we obtain $m(r_{0}) = m(V_{0})$ and $q(r_{0}) = q(V_{0})$, and hence
\begin{align}
M + \Delta M = \frac{1}{2} \left(r_{0} + \frac{(Q_{\text{ini}} + \Delta Q)^2}{r_{0}} \right).
\end{align}
Thus, the wormhole throat can be expressed by using $\Delta M$ as 
\begin{equation}
  r_0 = M + \Delta M \pm \sqrt{(M + \Delta M )^2 - (Q_{\text{ini}} + \Delta Q)^2}. 
\end{equation}
This is a generalization of the previous result, Eq.~\eqref{thrrad}, where $Q$ is replaced with $Q_{\text{ini}} + \Delta Q$.
Unlike the constant charge case, we do not necessarily have to choose $+$ sign here. For the throat radius to become real and positive, $\Delta M$ must satisfy $M + \Delta M > |Q_{\mathrm{ini}}+\Delta Q|$. We further demand $r_0 > |Q_{\mathrm{fin}}|$ and $r_- < r_0 < r_+$, where $r _{\pm}$ are outer and inner horizons of initial black hole, $r_{\pm} = M \pm \sqrt{M^2 - Q_{\mathrm{ini}}^2}$.
If we choose $r_0$ with $-$ sign, each condition becomes
\begin{align}
  |Q_{\mathrm{fin}}| < r_{0} & \Leftrightarrow ~ |Q_{\mathrm{fin}}| - M  < \Delta M < \frac{Q_{\mathrm{fin}}^2 + (Q_{\mathrm{ini}} + \Delta Q)^2}{2 | Q_{\mathrm{fin}}|} - M ,  \label{Qf<r0m}\\
  r_- < r_{0} & \Leftrightarrow ~ -\sqrt{M^2 - Q_{\mathrm{ini}}^2} < \Delta M < \frac{(Q_{\mathrm{ini}} + \Delta Q)^2 - Q_{\mathrm{ini}}^2}{2 \qty(M - \sqrt{M^2 - Q_{\mathrm{ini}}^2})}
  , \\
  r_0 < r_+ &  \Leftrightarrow ~ \Delta M < \sqrt{M^2 - Q_{\text{ini}}^2} ~ \text{or} ~ \frac{(Q_{\text{ini}}+\Delta Q)^2 - Q_{\text{ini}}^2}{2 \left(M + \sqrt{M^2 - Q_{\text{ini}}^2}\right)}< \Delta M. \label{r0m<rp}
\end{align}
Eq.~\eqref{Qf<r0m} can be satisfied only when $|Q_{\text{fin}}| < |Q_{\text{ini}} + \Delta Q|$.
When the first inequality in Eq.~\eqref{r0m<rp} is satisfied, the constraints can be summarized as 
\begin{align}
& \max \left\{  |Q_{\text{fin}}| - M, -\sqrt{M^2 - Q_{\text{ini}}^2}  \right\} < \Delta M
\notag\\
& < \min \left\{ 
\frac{Q_{\mathrm{fin}}^2 + (Q_{\mathrm{ini}} + \Delta Q)^2}{2 | Q_{\mathrm{fin}}|} - M,
\frac{(Q_{\mathrm{ini}} + \Delta Q)^2 - Q_{\mathrm{ini}}^2}{2 \qty(M - \sqrt{M^2 - Q_{\mathrm{ini}}^2})},
\sqrt{M^2 - Q_{\text{ini}}^2}
\right\}.
\end{align}
When the second inequality in Eq.~\eqref{r0m<rp} is satisfied, the parameters must satisfy $|Q_{\text{ini}}| < |Q_{\text{ini}}+\Delta Q|$. In this case $\Delta M > 0$ and the constraints can be summarized as 
\begin{align}
&\max\left\{ 
|Q_{\mathrm{fin}}| - M, \frac{(Q_{\text{ini}}+\Delta Q)^2 - Q_{\text{ini}}^2}{2 \left(M + \sqrt{M^2 - Q_{\text{ini}}^2}\right)} \right\}
< \Delta M
\notag\\
& <  
\min \left\{
\frac{Q_{\mathrm{fin}}^2 + (Q_{\mathrm{ini}} + \Delta Q)^2}{2 | Q_{\mathrm{fin}}|} - M, 
\frac{(Q_{\mathrm{ini}} + \Delta Q)^2 - Q_{\mathrm{ini}}^2}{2 \qty(M - \sqrt{M^2 - Q_{\mathrm{ini}}^2})}
\right\}.
\end{align}

 On the other hand, $r_0$ with $+$ sign case, each condition becomes
\begin{align}
  |Q_{\mathrm{fin}}| < r_{0} & \Leftrightarrow ~ \min \qty{|Q_{\mathrm{fin}}| - M, \frac{Q_{\mathrm{fin}}^2 + (Q_{\mathrm{ini}} + \Delta Q)^2}{2 | Q_{\mathrm{fin}}|}-M} < \Delta M,  \\
  r_- < r_0 & \Leftrightarrow ~ \min \qty{- \sqrt{M^2 - Q_{\mathrm{ini}}^2}, \frac{(Q_{\mathrm{ini}} + \Delta Q)^2 - Q_{\mathrm{ini}}^2}{2 \qty(M - \sqrt{M^2 - Q_{\mathrm{ini}}^2})}} < \Delta M, \\
  r_0 < r_+ &  \Leftrightarrow ~ \Delta M < \min \qty{\sqrt{M^2 - Q_{\mathrm{ini}}^2},\frac{(Q_{\mathrm{ini}} + \Delta Q)^2 - Q_{\mathrm{ini}}^2}{2 \qty(M + \sqrt{M^2 - Q_{\mathrm{ini}}^2})}}. 
\end{align}
Putting them together, we obtain
\begin{align}
    &\max \Bigg\{\min \qty{|Q_{\mathrm{fin}}| - M, \frac{Q_{\mathrm{fin}}^2 + (Q_{\mathrm{ini}}+\Delta Q)^2}{2 | Q_{\mathrm{fin}}|} - M},\\
    & \qquad \qquad \qquad \min\qty{- \sqrt{M^2 - Q_{\mathrm{ini}}^2}, \frac{(Q_{\mathrm{ini}} + \Delta Q)^2 - Q_{\mathrm{ini}}^2}{2 \qty(M - \sqrt{M^2 - Q_{\mathrm{ini}}^2})}} \Bigg\}\notag\\
    & < \Delta M < \min \qty{\frac{(Q_{\mathrm{ini}}+\Delta Q)^2 - Q_{\mathrm{ini}}^2}{2 \qty(M + \sqrt{M^2 - Q_{\mathrm{ini}}^2})}, \sqrt{M^2 - Q_{\mathrm{ini}}^2}}.
\end{align}

Let us focus on the case with $+$ sign of $r_{0}$. If we assume $ |Q_{\mathrm{ini}} + \Delta Q| > |Q_{\mathrm{ini}}|$, $\Delta M$ could be positive. This is a sharp contrast to the neutral shell cases. We remark that this does not contradict the violation of the null energy condition: The maximum value of the surface energy density sandwiched between the black hole and the Vaidya region, $\sigma_{1}$ in Eq.~\eqref{sigma1}, reduces to
\begin{equation}
  \begin{aligned}
    \sigma_{1} &= \frac{1}{8 \pi r^2} \qty(2 \Delta M - \frac{1}{r} ((Q_{\mathrm{ini}} + \Delta Q)^2 - Q^2_{\mathrm{ini}})) \\
    &< \frac{(Q_{\mathrm{ini}} + \Delta Q)^2 - Q^2_{\mathrm{ini}}}{8 \pi r^2} \qty(\frac{1}{r_+} - \frac{1}{r} ) < 0,
  \end{aligned}
\end{equation}
inside the outer horizon $r < r_+$.  Thus, at the point of wormhole formation, this surface energy indeed has a negative energy.

Finally, continuity of the energy-momentum tensor and the electric charge is investigated in App.~\ref{sec:app C}. The expressions for $\Delta M$ and $\Delta Q$ are given in Eqs.~\eqref{eq:continuity of T} and \eqref{eq:continuity of j}, respectively.

\section{Conclusions and discussions} \label{sumdis}

In this work, we constructed static, spherically symmetric, charged, traversable wormhole solutions supported by bidirectional null dust with negative energy. 
By solving the Einstein--Maxwell equations, we obtained wormhole geometries that satisfy essential criteria such as the existence of a throat, the flare-out condition, finite proper distance, and regular curvature at the throat. Numerical solutions are provided in Sec.~\ref{exthra}. There, we found that a maximal value of the charge $Q$ exists in the series of symmetric wormhole solutions. We also found that, with this class, the maximum throat size is achieved by Hayward's chargeless wormhole solution \cite{Hayward:2002pm}.  
However, we also found that our solutions, as well as Hayward's chargeless solutions, are not asymptotically flat, but instead possess curvature singularities at infinite areal radius $r \to \infty$.
This behavior originates from the nature of the negative energy source used to support the wormhole, which continues to dominate even at large distances in this setup.
The violation of asymptotic flatness in these solutions stands in contrast to the solution obtained by Maldacena~\cite{Maldacena:2018gjk}, where the Casimir energy is confined to a finite region. 
This suggests that our solutions should be interpreted as describing a finite portion of a physically reasonable asymptotically flat wormhole, 
by assuming that the negative energy in the form of bidirectional null dust is turned off at sufficiently large distances.

We also discussed a dynamical formation process of our wormhole geometry from a RN black hole spacetime as an extension of Ref.~\cite{Koyama:2004uh}. 
By introducing impulsive null shells followed by a continuous stream of neutral, negative energy null dust modeled by, Vaidya spacetimes, we demonstrated that a black hole can transition into a traversable wormhole. 
The junction conditions for the energy-momentum tensor of the shell were carefully analyzed, and we found the consistency relations among the initial mass, charge, and injected shell energy. 
The throat radius 
was explicitly determined by three physical parameters: the black hole mass $M$, electric charge $Q$, and shell energy $\Delta M$.
Thus, the structure of the wormhole was also determined by these parameters.

We further generalized this setup to the case where the electric charge varies across spacetime by allowing the null dust to carry charge. 
This scenario enabled the construction of wormholes whose charge differs from that of the initial black hole. 
We found that the throat radius of the created wormhole is determined by the black hole mass $M$, electric charge $Q$, shell's energy measured in the past null infinity $\Delta M$ and shell's charge $\Delta Q$. 
We found that, in contrast to the case of neutral null dust, the formation remains viable even with positive $\Delta M$, as long as the local null energy condition is violated due to the electric self-energy contribution. 

Future work could extend these results by exploring: 
\begin{itemize}
  \item Realization of the dynamical formation process by physically realistic energy sources such as full quantum field stress tensors,
  \item Solutions with reduced symmetry (e.g., axisymmetric ~\cite{Makita:2025bao}, stationary~\cite{Teo:1998dp,Kashargin:2007mm,Kashargin:2008pk,Volkov:2021blw} or self-similar wormholes), 
  \item Detailed analysis of the causal structure and stability of the constructed geometries,
  \item Models with regularized asymptotic singularities beyond the null dust or symmetric assumption, for example, in order to describe the entire structure of wormhole with Casimir energy~\cite{Maldacena:2018gjk},
  \item Realization of dynamical formation scenarios consistent with energy conditions, e.g., through theories beyond Einstein gravity~\cite{Kanti:2011jz, Jusufi:2020yus, Mehdizadeh:2018smu, Moraes:2016akv, Godani:2018blx}.
\end{itemize}

Overall, this work contributes to the broader understanding of how semiclassical effects can mediate nontrivial spacetime topologies and bridges the gap between theoretical wormhole constructions and dynamical black hole physics.

\section*{Acknowledgments}

This work was partially supported by Grants-in-Aid for Scientific Research from the Japan Society for the Promotion of Science (JSPS) and the Ministry of Education, Culture, Sports, Science and Technology (MEXT) of Japan
under Grant Numbers
JP23H01170~(YK),
JP24KJ1223~(DS), 
JP21H05189~(DY).
This work was also financially supported by JST~SPRING, Grant Number 
JPMJSP2125. The author (KU) would like to take this opportunity to thank the 
“THERS Make New Standards Program for the Next Generation Researchers.” 

\appendix

\section{The stress tensor of electromagnetic field} \label{emtemf}

We will review the energy-momentum tensor of an electromagnetic field in several kinds of geometries. 
We only consider a spherical symmetric case, where the metric has the general form,
\begin{equation}
  ds^2 = - e^{2 \psi(t, r)} f(t,r) dt^2 + \frac{1}{f(t,r)} dr^2 + r^2 d \Omega^2. 
\end{equation}
We assume there are no sources or currents for the Maxwell equations.
Then the electromagnetic field should satisfy,
\begin{equation}
  \tensor[]{F}{^\mu ^\nu _; _\nu} = \frac{1}{\sqrt{-g}} (\sqrt{-g} F^{\mu \nu})_{, \nu} = 0. 
\end{equation}
Here $F^{\alpha \beta}$ is the field strength. 
Using the field strength, the energy-momentum tensor can be described as
\begin{equation}
  \tensor[]{T}{^\mu_\nu} = \frac{1}{4 \pi} \qty(F^{\mu \alpha} F_{\nu \alpha} - \frac{1}{4} \tensor[]{\delta}{^\mu _\nu} F^{\alpha \beta} F_{\alpha \beta}). \label{eleene}
\end{equation}

\subsection{Static spacetime}

First, we consider the static spacetime, i.e. the metric functions $\psi(r)$ and $f(r)$ do not depend on time $t$.
In this case, from Maxwell equation and symmetry, the field strength also becomes static and spherically symmetric. 
The non-vanishing component of the Maxwell equations is
\begin{equation}
  \begin{aligned}
    \dv{r} \qty( e^{\psi(r)} r^2 F^{t r}) = 0, \\
    \therefore F^{t r} = e^{-\psi(r)} ~ \frac{Q}{r^2}. \label{fields}
  \end{aligned}
\end{equation}
We only considered the electric field here. 
The $Q$ is a constant of integration which we can interpret as electric charge. 
Substituting this to Eq.~\eqref{eleene}, we get the energy-momentum tensor as 
\begin{equation}
  \tensor[]{T}{^\mu_\nu} = \frac{Q^2} {8 \pi r^4} \mathrm{diag} (-1, -1, 1, 1). \label{stress}
\end{equation}
We can use this form not only in black hole solutions but also in static wormhole geometry in Sec.~\ref{chawhs}. 
We remark that this matrix form of energy-momentum tensor is also true in Eddington-Finkelstein coordinate and double null coordinate, and even other coordinates that do not mix $\{t,r\}$ and $\{\theta, \phi\}$, because $t, r$ components and $\theta, \phi$ components are direct sum and respectively proportional to unit matrix.

\subsection{Vaidya spacetime}

The Vaidya spacetime is not static, so the above results cannot be applied immediately. 
Nonetheless, we can show that the results are also true in Vaidya geometry. 
The line element can be written in Eddington-Finkelstein coordinates,
\begin{equation}
  ds^2 = - f(v, r) dv^2 + 2 dv \, dr + r^2 d\Omega^2. 
\end{equation}
The metric determinant $\sqrt{-g} = r^2 \sin \theta$ does not depend on $v$, so again the field strength becomes static. 
Thus, in the same way as Eq.~\eqref{fields}, the field strength becomes $F^{vr} = Q / r^2$ and we again get Eq.~\eqref{stress}.

\section{Junction condition of electromagnetic field} \label{elmgjc}

In this section we derive the junction condition of electromagnetic fields based on Maxwell equation. 
We follow the procedure of the derivation of geometries' junction condition developed in Refs.~\cite{Poisson:2009pwt, Barrabes:1991ng}. 

We will use the same notation shown in Sec.~\ref{juncon}. 
We express the electromagnetic field strength as
\begin{equation}
  F^{\mu \nu} = F_{(+)}^{\mu \nu} \Theta (\Phi) + F_{(-)}^{\mu \nu} \Theta (- \Phi), 
\end{equation}
where $\Theta (\Phi)$ is the Heaviside distribution. 
We substituting this into Maxwell equation $\nabla_\mu F^{\mu\nu} = j^\nu$.
By using $\partial_\mu \Theta (\Phi) = \partial_\mu \Phi ~ \delta(\Phi) = - \alpha n_\mu \delta(\Phi)$, the left hand side becomes
\begin{equation}
  \nabla_\mu F_{(+)}^{\mu \nu} \Theta (\Phi) + \nabla_\mu F_{(-)}^{\mu \nu} \Theta (- \Phi) - \alpha [F^{\mu \nu}] n_\mu \delta(\Phi). 
\end{equation}
The third term is identified to the surface current, which gives the junction condition. 
\begin{align}
  j^\mu_{\mathrm{(surface)}} &= - \alpha ~i^\mu ~\delta(\Phi), \\
  [F^{\mu \nu}] n_\mu &= i^\nu. 
\end{align}
We can rewrite this in coordinate invariant form;
\begin{align}
  \qty[F_{\mu \nu}] n^{\mu} e^\nu_{(A)} &= i_\nu e^\nu_{(A)}, \\
  \qty[F_{\mu \nu}] n^{\mu} N^\nu &= i_\nu N^\nu, \\
  0 &= i_\nu n^\nu. 
\end{align}

\section{Continuity across the intersection between $\Sigma_{1}$ and $\Sigma_{2}$}
\label{sec:app C}
In the main part of this paper, we mainly focus on the junction in the right region.
Here we briefly discuss the junction between Vaidya and wormhole in the left region and clarify the continuity conditions for the energy-momentum tensor and electric current localized on $\Sigma_{2}$ beyond the intersection with $\Sigma_{1}$.

Due to the flipping symmetry of the system, analysis in the left region is essentially same as that in the right region.
Only the difference is we use the coordinate $U$ or $u$ for the \textit{in-going} Eddington--Finkelstein coordinates instead of $V$ or $v$.
Thus, the metric of Vaidya and wormhole region can be expressed as
\begin{align}
  &\text{Vaidya} & : & 
  \begin{cases}
    ds^2 = - e^{\psi_V} dU (f_V e^{\psi_V} dU - 2 dr) + r^2 d \Omega^2, \\
    f_V = 1 - \frac{2 m(U)}{r} + \frac{q(U)^2}{r^2} , \qquad \psi_V = 0, \label{chvaf2}
  \end{cases} \\
  &\text{Wormhole} & : &
  \begin{cases}
    ds^2 = - e^{\psi_W} dv (f_W e^{\psi_W} dv + 2 dr) + r^2 d \Omega^2, \\
    f_W = 1 + \frac{Q_{\mathrm{fin}}^2}{r^2} - \frac{B(r)}{r}, \\
    e^{\psi_W} = \sqrt{-\frac{2 \lambda}{B'(r) (1 + Q^2_{\mathrm{fin}} / r^2 - B / r)}}. \label{chwof2}
  \end{cases} 
\end{align}
The location of $\Sigma_{1}$ is expressed as $U = U_{1}(r)$ in the Vaidya region and $v = v_{0}$ in the wormhole region.
The transverse vector $N_{\mu}$ across $\Sigma_{2}$ in the Vaidya-Wormhole junction can be expressed as 
\begin{align}
    \tilde{N}_{\mu} dx^{\mu} = - \frac{f_{V}}{2} e^{\psi_{V}} dU
\end{align}
in the Vaidya region and
\begin{align}
    N_{\mu} dx^{\mu} = - \frac{f_{W}}{2} e^{\psi_{W}} d v - d r,
\end{align}
in the wormhole region.

We can introduce global double null coordinates $(u,v)$ for wormhole region by introducing another null coordinate $u$ through
\begin{align}
    u = v - 2 r_{*},
\end{align}
which corresponds to the transformation of the coordinate basis,
\begin{align}
    d u = d v - 2 d r_{*} = d v + \frac{2}{f_{W} e^{\psi_{W}}} d r.
\end{align}
Here we use the fact that $r(r_{*})$ is decreasing function in the left region. By using $u$ coordinate, we can express the transverse vector $N_{\mu}$ as 
\begin{align}
 N_{\mu} dx^{\mu} = - \frac{f_{W}}{2} e^{\psi_W} d u.
\end{align}
Since $\tilde{N}_{\mu}$ and $N_{\mu}$ are identified at the junction across $\Sigma_{1}$, we obtain
\begin{align}
    f_{V} e^{\psi_{V}} d U|_{\Sigma_{1}} = f_{W} e^{\psi_{W}} d u |_{\Sigma_{1}}, \label{eq:du = dU}
\end{align}
which provides the relation between $dU$ and $du$ at $\Sigma_{1}$. 

As an application of Eq.~\eqref{eq:du = dU}, we can relate the energy-momentum tensors on $\Sigma_{2}$ across the intersection with $\Sigma_{1}$.
In the Black hole-Vaidya junction across $\Sigma_{2}$, thus, in the left region, the energy-momentum tensor on the shell is given by
\begin{align}
    T_{1}{}_{\mu\nu} dx^{\mu} dx^{\nu} & = \sigma_{1} n_{\mu} dx^{\mu} ( e^{- \psi_{V}} n_{\nu} dx^{\nu} \delta(U - U_{0}) ) \\
    &= \frac{\sigma_{1}}{f_{V}} f_{V} e^{\psi_{V}} d U d \Theta(U - U_{0}).
\end{align}
Here we use the expression 
\begin{align}
n_{\mu} dx^{\mu} = - e^{\psi_{V}} d U.
\end{align}
On the other hand, in the Vaidya-Wormhole junction across $\Sigma_{2}$, the energy-momentum tensor on the shell is
\begin{align}
T_{2}{}_{\mu\nu} dx^{\mu} dx^{\nu} &= \sigma_{2} n^{\mu} dx^{\mu} \left( e^{-\psi_{W}} n^{\nu} dx^{\nu} \delta(u - u_{0}) \right) \\
&= \frac{\sigma_{2}}{f_{W}} f_{W} e^{\psi_{W}} du d\Theta(u - u_{0}).
\end{align}
Then, from Eq.~\eqref{eq:du = dU}, the requirement of the continuity of the energy-momentum tensor across $\Sigma_{1}$, we obtain,
\begin{align}
\lim_{r \to r_{0}} T_{1}{}_{\mu\nu} dx^{\mu} dx^{\nu} = \lim_{r \to r_{0}} T_{2}{}_{\mu\nu} dx^{\mu} dx^{\nu}  
\Leftrightarrow
\lim_{r\to r_{0}} \sigma_{1} = \lim_{r \to r_{0}} \frac{f_{V}}{f_{W} e^{\psi_{W}}} e^{\psi_{V}}\sigma_{2}.
\end{align}
Note that we use the fact that the step function $\Theta$ is independent of how to express the surface $\Sigma_{2}$.

One can show that the combination $f_{V}/(f_{W}e^{\psi_{W}})$ is finite in the limit $r \to r_{0}$ directly plugging the expressions obtained in the main section, noting that $f_{V}$ can be expressed as 
\begin{align}
    f_{V} & = f_{W} - \frac{2 C(r)}{r} \mathrm{e}^{- \psi_{W}}.
\end{align}
Explicitly, we obtain
\begin{align}
    \lim_{r \to r_{0}} \frac{f_{V}}{f_{W} e^{\psi_{W}}} &= \lim_{r \to r_{0}} \frac{f_{W} - \frac{2 C(r)}{r} \mathrm{e}^{- \psi_{W}}}{f_{W} \mathrm{e}^{\psi_{W}}} \\
    &= \lim_{r \to r_{0}} \frac{f_{W} \mathrm{e}^{\psi_{W}} - \frac{2 C(r)}{r}}{f_{W} \mathrm{e}^{2\psi_{W}}} \\
    &= \frac{B'(r_{0})}{2 \lambda} \frac{2 C(r_{0})}{r_{0}} .
\end{align}
Also, using the expression
\begin{align}
\lim_{r\to r_{0}}\sigma_{1} &= \frac{\Delta M}{4 \pi r_{0}^2} -\frac{(Q_{\mathrm{ini}} + \Delta Q)^2 - Q_{\mathrm{ini}}^2}{8 \pi r_0^3}, \\
\lim_{r \to r_{0}} e^{\psi_{W}} \sigma_{2} &= \frac{C(r_{0})}{4 \pi r_{0}^2}.
\end{align}
Combining the above results, the continuity condition can be expressed as
\begin{align}
    \Delta M = - \left( \frac{- B'(r_{0})}{\lambda} \right) \frac{C(r_{0})^2}{r_{0}} +\frac{(Q_{\text{ini}} + \Delta Q)^2 - Q_{\mathrm{ini}}^2}{2 r_0}. \label{eq:continuity of T}
\end{align}
An immediate consequence is the negativity of $\Delta M$ when the electric charge on the shell $\Delta Q$ vanishes, because $B'(r_{0})$ is negative.

Similarly, the electric flux localized on $\Sigma_{2}$ can be evaluated as  
\begin{align}
 j_{1}{}_{\mu} dx^{\mu} =  - i_{1}^{\nu} N_{\nu}  d \Theta(U - U_0), 
\end{align}
for the black hole - Vaidya junction, and 
\begin{align}
 j_{2}{}_{\mu} dx^{\mu} = - i_{2}^{\nu} N_{\nu} d \Theta(u - u_{0}) ,
\end{align}
for the Vaidya - Wormhole junction. 
Hence, the continuity across the intersection with $\Sigma_{1}$ can be expressed as 
\begin{align}
 \lim_{r \to r_{0}} i^{\mu}_{1} N_{\mu} =  \lim_{r \to r_{0}} i_{2}^{\mu} N_{\mu},
\end{align}
which reduces to
\begin{align}
    \Delta Q = \frac{Q_{\text{fin}} - Q_{\text{ini}}}{2}. \label{eq:continuity of j}
\end{align}

\bibliography{thurs} 
\bibliographystyle{JHEP}

\end{document}